\documentclass[article]{aa}
\usepackage{graphicx}
\usepackage{amsmath}
\usepackage{pstricks}
\begin{document}
\title{Deconvolution of HST images of the Cloverleaf gravitational lens
\subtitle{Detection of the lensing galaxy and a partial Einstein ring}
\thanks{Based on observations made with the NASA/ESA HST Hubble Space Telescope, obtained from the data archive at the Space Science Institute. STScI is operated by AURA, the Association of Universities for Research in Astronomy, Inc., under NASA contract NAS- 5-26555.}}

\author{V. Chantry 
\thanks{Research Fellow, Belgian National Fund for Scientific Research (FNRS)}
\and P. Magain 
}
\offprints{Virginie Chantry (Virginie.Chantry@ulg.ac.be)}
\institute{Institut d'Astrophysique et de G\' eophysique, Universit\' e de
  Li\`ege, All\'ee du 6 Ao\^ut, 17, Sart Tilman (Bat. B5C), Li\`ege 1, Belgium
}

\date{}

\abstract{Archival HST/NICMOS--2 images of the {\em Cloverleaf} gravitational lens (H1413+117), a quadruply imaged quasar, have been analysed with a new method derived from the MCS deconvolution algorithm (Magain et al., \cite{MCS}). This method is based on an iterative process which simultaneously allows to determine the Point Spread Function (PSF) and to perform a deconvolution of images containing several point sources plus extended structures.  As such, it is well adapted to the processing of gravitational lens images, especially in the case of multiply imaged quasars.
Two sets of data have been analysed : the first one, which has been obtained through the F160W filter in 1997, basically corresponds to a continuum image, while the second one, obtained through the narrower F180M filter in 2003, is centered around the forbidden [O{\sc III}] emission lines at the source redshift, thus probing the narrow--line region of the quasar. 
The deconvolution gives astrometric and photometric measurements in both filters and reveals the primary lensing galaxy as well as a partial Einstein ring. The high accuracy of the results is particularly important in order to model the lensing system and to reconstruct the source undergoing the strong lensing. The reliability of the method is checked on a synthetic image similar to H1413+117.
}

\titlerunning{Deconvolution of HST images of the Cloverleaf gravitational lens}

\maketitle

\section{Introduction}
Four years after its discovery by Hazard et al. (\cite{hazard}), the quasi--stellar object (QSO) H1413+117 was identified as a gravitational lens  by Magain et al.\  (\cite{magain_a}). This system, consisting in 4 components of comparable brightness separated by $\sim 1$ arcsec, is best known as the {\em Cloverleaf}.  It is also one of the brightest quasars amongst the BAL (Broad Absorption Line) class, with a redshift of 2.558 and an apparent visual magnitude of 17.  The lensing galaxy was detected by Kneib et al. (\cite{kneib}) from a careful PSF subtraction on near--infrared Hubble Space Telescope (HST) images.

The aim of the present paper is to present a method which simultaneously allows to perform PSF determination and deconvolution on images containing several point sources superimposed on a diffuse background, and to apply it to HST images of the Cloverleaf gravitational lens.  We show that this method  permits a more accurate astrometry of the system and a better characterisation of the lensing galaxy. Moreover, it also allows the detection of additional structures, such as parts of an Einstein ring.

This method is based on the MCS deconvolution algorithm (Magain, Courbin \& Sohy, \cite{MCS}) which, unlike most deconvolution methods, ensures that the deconvolved image, which has a well defined Point Spread Function (PSF), conforms to the sampling theorem.  The method also leads to a decomposition of the light distribution into a sum of point sources (of known shape) and a diffuse background.

More recently, Magain et al. (\cite{psfsimult}) presented a method, derived from MCS, to determine the PSF on images consisting of (possibly blended) point sources.  This method works well, even in very crowded fields, when no point source is sufficiently isolated to derive an accurate PSF from standard techniques.

The algorithm presented here extends the method of Magain et al. (\cite{psfsimult}) to images containing a mixture of point sources and diffuse background. It is based on an iterative scheme, in which both the PSF and the diffuse background are improved step by step.

In section \ref{image} we describe the input data and their reduction. The method used to obtain both the PSF and the deconvolved images is described in section \ref{method}.  The results are presented and discussed in section \ref{results}. The accuracy of our results is tested by applying the method to a synthetic image with the same basic configuration as the Cloverleaf (see section  \ref{synthetic}). Finally we conclude in section \ref{conclusion}.

\section{HST Images}
\label{image}

\begin{figure}[h]
\centering
\includegraphics[scale=0.25]{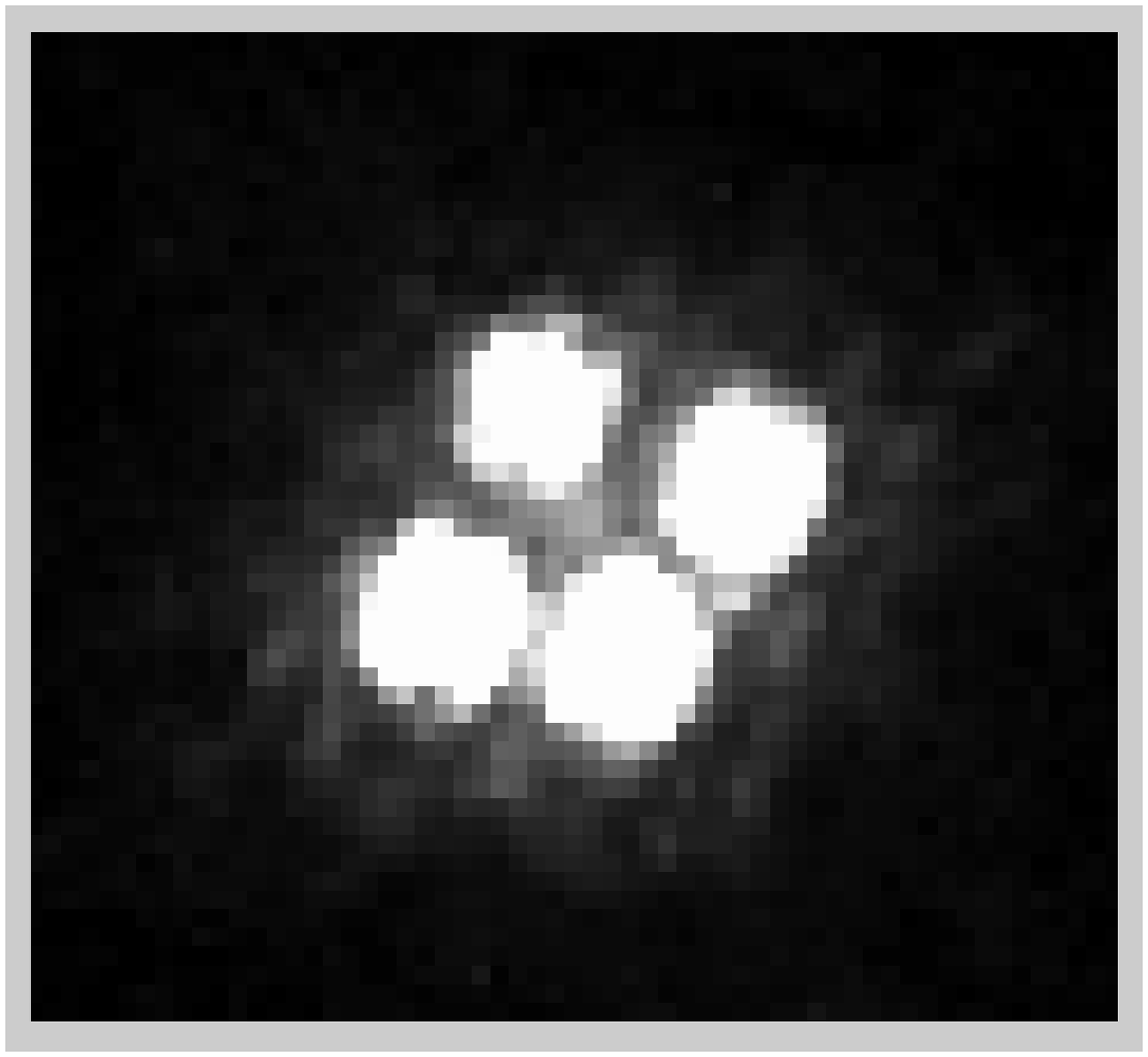}
\includegraphics[scale=0.25]{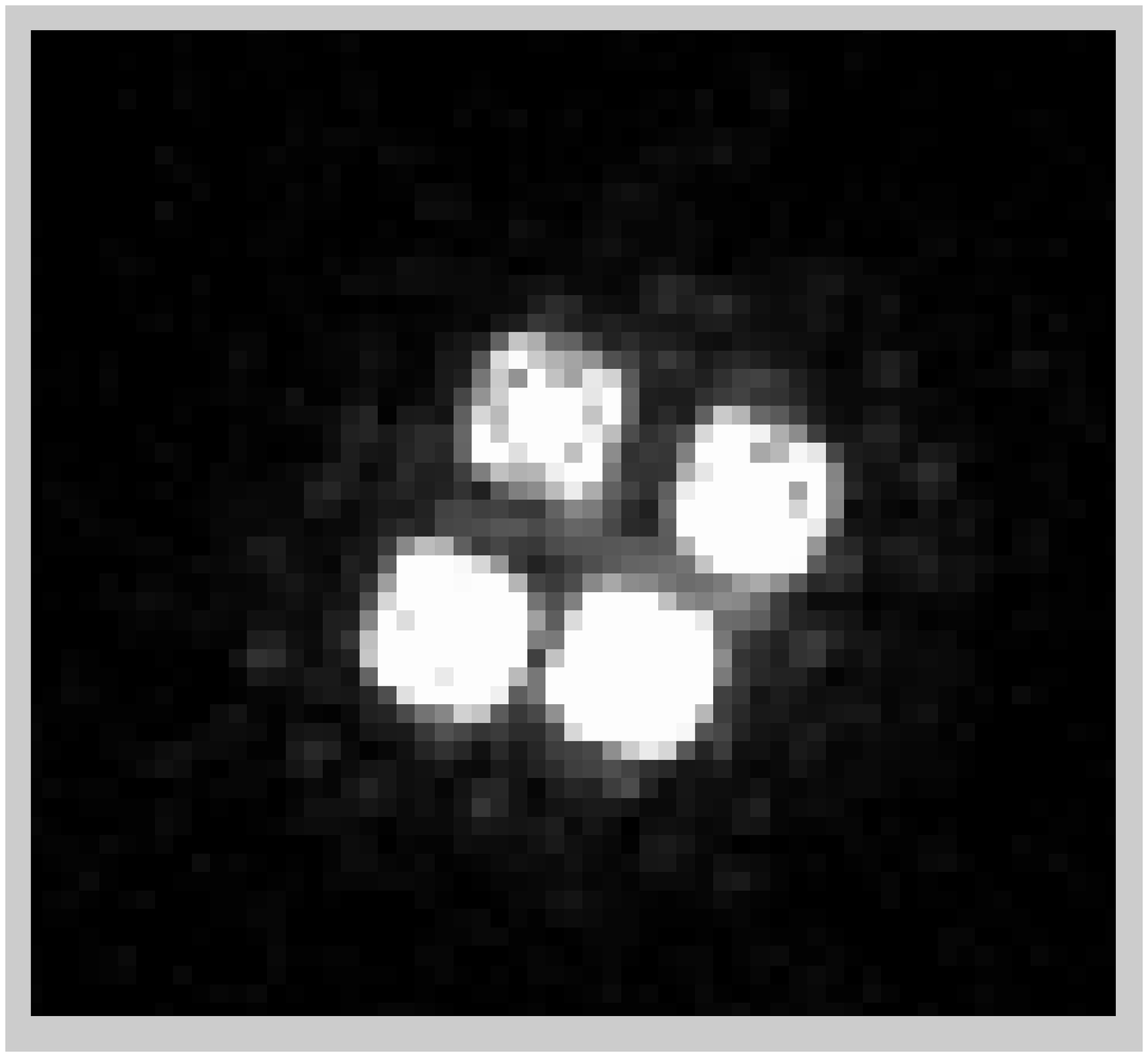}
\caption{{\it  Left} : combination of the 4 calibrated images from the F160W filter data set obtained with the HST/NICMOS--2, the grey scale going from 0\% (black) to 3.2\% (white) of the maximum intensity; {\it Right} : combination of the 8 calibrated images from the F180M filter data set obtained with the HST/NICMOS--2, the grey scale going from 0\% (black) to 4.7\% (white) of the maximum intensity. 
The structure of the PSF is obvious. North is to the top and East to the left.  }
\label{Nicmos}
\end{figure}

The first set of HST data was obtained on the $28^{th}$ of December 1997 by the camera 2 of NICMOS (Near Infrared Camera and Multi--Object Spectrometer) with the F160W filter (wide band filter), corresponding approximately to the near--IR H--Band (PI: E. Falco). We used the 4 calibrated images, i.e. treated by the HST image reduction pipeline (CALNICA). Each of them has an exposure time of 639.9389 s and a mean pixel size of 0\farcs07510 according to Tiny Tim\footnote{Tiny Tim is a software package which allows to generate the HST Point Spread Function for each instrument, filter and observing configuration.}
software v 6.3 (Kris \& Hook, \cite{tinytim}). These images were obtained in the MULTIACCUM mode : each of them is a combination of several samples (19 in the present case). A combination of these 4 images is shown on the left panel of Fig. \ref{Nicmos}.

The second set of images was obtained on the $10^{th}$ of July 2003 with the same instrument and the F180M medium band filter (PI: D. A. Turnshek). As for the F160W filter we used the calibrated images, here 8 images, 4 of them being a combination of 18 samples and the other 4 being a combination of 16 samples. The first 4 have an exposure time of 575.9418 s and the latter 4 an exposure time of 447.9474 s. The mean pixel size is, again according to Tiny Tim software, 0\farcs07568. A combination of these calibrated images is shown on the right panel of Fig. \ref{Nicmos}.

The wavelength ranges of these two filters are partly superimposed : the passband of the F160W filter is $1.4 \ \mu m \leq \lambda \leq 1.8 \ \mu  m$ while it amounts to $1.76 \ \mu m \leq 1.83 \ \mu m$ for the F180M filter. The latter was chosen in order to include the oxygen [O{\sc III}] forbidden line doublet (499 -- 501 nm) at the redshift of the QSO.

The image reduction is divided in two parts : the image cleaning and the calculation of the sigma images (i.e. images containing the standard deviations of the pixel intensities). The first step of the first part consists in computing the intensities in counts per pixel. The second step consists in removing the sky background. As the NIC--2 detector is composed of 4 quadrants, it is necessary to subtract a different constant value for each of them. These constants were derived from the parts of the image where there is no obvious light source. 

The second step consists in the calculation of the sigma images. We start from the sigmas calculated by the pipeline CALNICA. Two effects are then corrected. First, we take into account the underevaluation of the standard deviation for the negative pixels (by replacing all negative intensities by a null value). Secondly, we make use of the HST flag files indicating bad pixels, e.g. cold or hot pixels. It allows us, using the inverted sigma images, to put the statistical weight of such bad pixels to zero so that the information they provide has no weight in the deconvolution.

Let us mention that we do not remove the cosmic rays impacts from the images during the reduction process. We use the deconvolution residuals (see below) to spot the pixels likely of having been contaminated by a cosmic ray. We then put the inverted sigma value of such pixels to zero.

All these manipulations are carried out with the IRAF\footnote{IRAF, Image Reduction and Analysis Facility, is distributed by the National Optical Astronomy Observatories, which are operated by AURA, the Association of  Universities for Research in Astronomy, Inc., under cooperative agreement with the NSF, National Science Foundation.} package.

\section{Method}
\label{method}

The same technique, based on the MCS deconvolution algorithm, has been applied to both sets of images in order to improve their resolution and sampling and, most importantly, to detect any significant extended structure which might be hidden by the complex PSFs. The method is based on the simultaneous deconvolution of all images from a set, as explained, e.g., in Courbin et al. (\cite{courbin}).  This means that we attempt to find a light distribution that is compatible with all images obtained in a given instrument configuration (e.g.  through a given filter). To do this we allow a spatial translation in between the individual images and, in some cases, a variation of the point source intensities. In order to improve the resolution while keeping a well sampled light distribution, we use a sampling step 2 times smaller than the original pixel size and we choose, as the final PSF (i.e. the PSF of the deconvolved images), a Gaussian with 2 pixels Full--Width--at--Half--Maximum (FWHM). Let us mention that, since the HST PSF varies with the position in the focal plane, and since the object is located in different parts of the detector at each exposure, each original image has its own individual PSF.

The originality of the present method is that the same images are used both to determine the PSF and to perform the deconvolution (basically to detect the diffuse background and to obtain the astrometry and photometry of all objects).  It works only if there are several point sources in the field : this allows to distinguish the structure belonging to the PSF (and thus appearing in the vicinity of each point source) from the diffuse background, assumed not to be identical around each source.

\begin{figure}
\centering
\includegraphics[width=4.2cm]{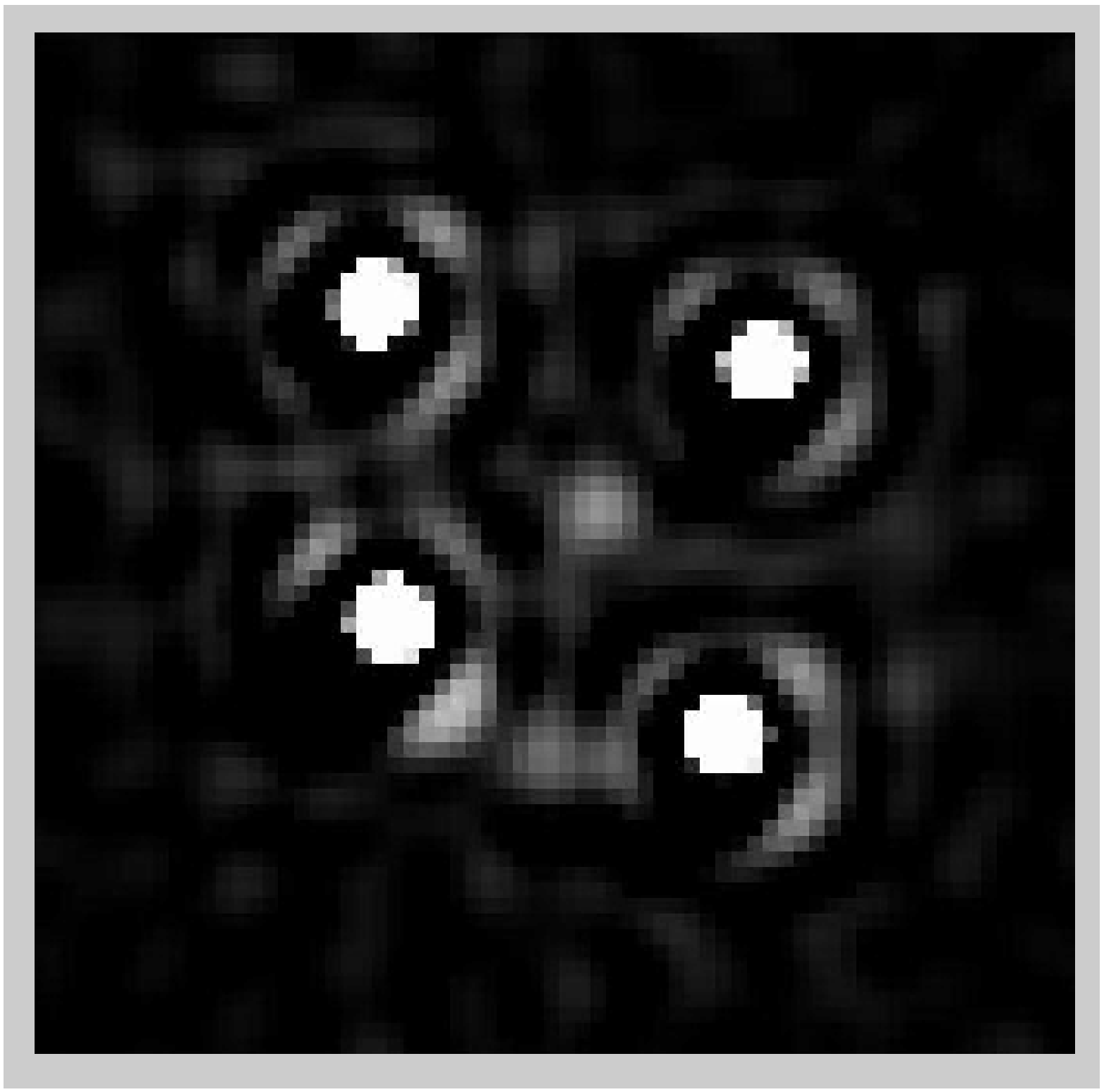}
\includegraphics[width=4.2cm]{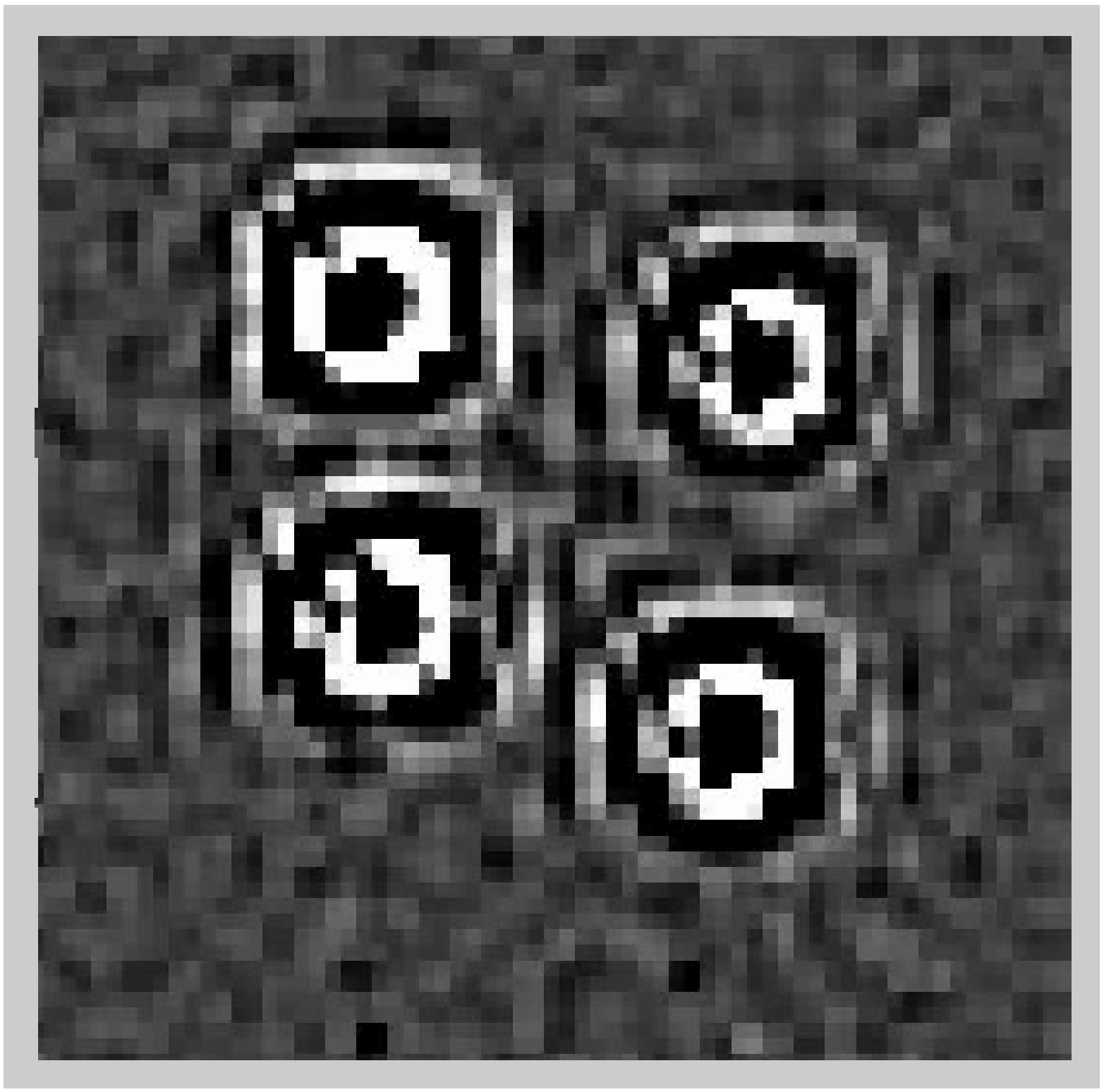}
\caption{Results of the simultaneous deconvolution for the F160W data set using a deconvolved Tiny Tim PSF. {\it Left} : deconvolved image, the grey scale going from 0\% (black) to 0.45\% (white) of the maximum intensity. {\it Right} : residuals (see text) of the deconvolution. The remnant structure around each point sources is obvious.}
\label{dec0_tiny}
\end{figure}

This new method is based on an iterative process. We start with a first approximation of the PSF constructed by the Tiny Tim software (see Fig. \ref{tinytim}) with a sampling step two times smaller than the original one. That PSF is deconvolved by the final Gaussian PSF in order to obtain the deconvolution kernel that we call the PSF $s_0(\vec{x})$.  This is a
reasonable first approximation, however not accurate enough to obtain trustworthy deconvolved images.  Indeed, when using that PSF for deconvolving the original images, that we call $d_0(\vec{x})$, significant structure appears around each point source, clearly showing that the Tiny Tim PSF departs from the actual one (see Fig. \ref{dec0_tiny}).  

Since no extra images of stars are available in the field to improve this PSF, we have to use the information in the point sources of the Cloverleaf itself. However, we know that there might be some extra structure under the 4 point sources, as well as a contribution from the lensing object. That is why we proceed as follows:

\begin{figure} [b]
\centering
\includegraphics[width=4.2cm]{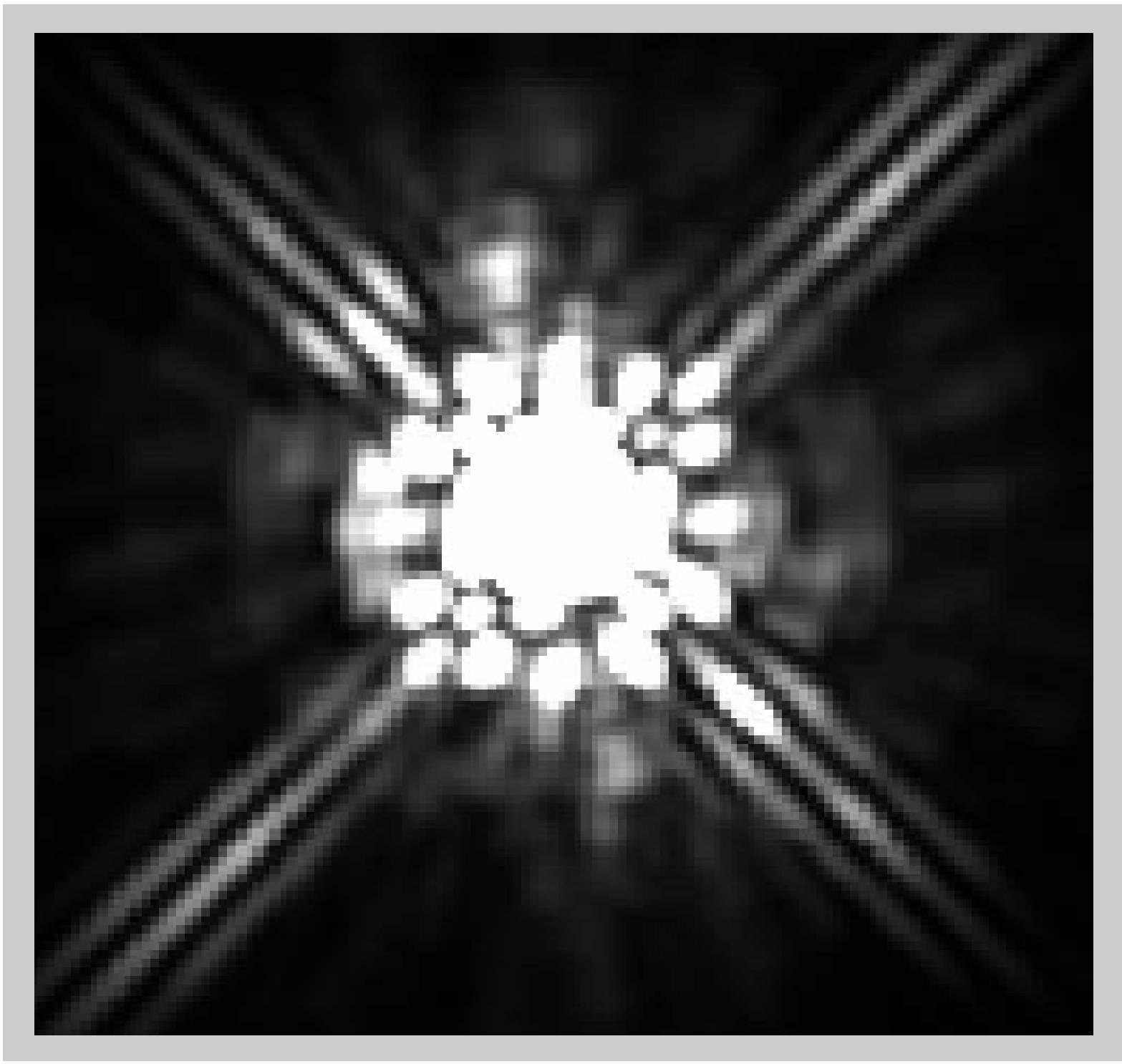}
\includegraphics[width=4.2cm]{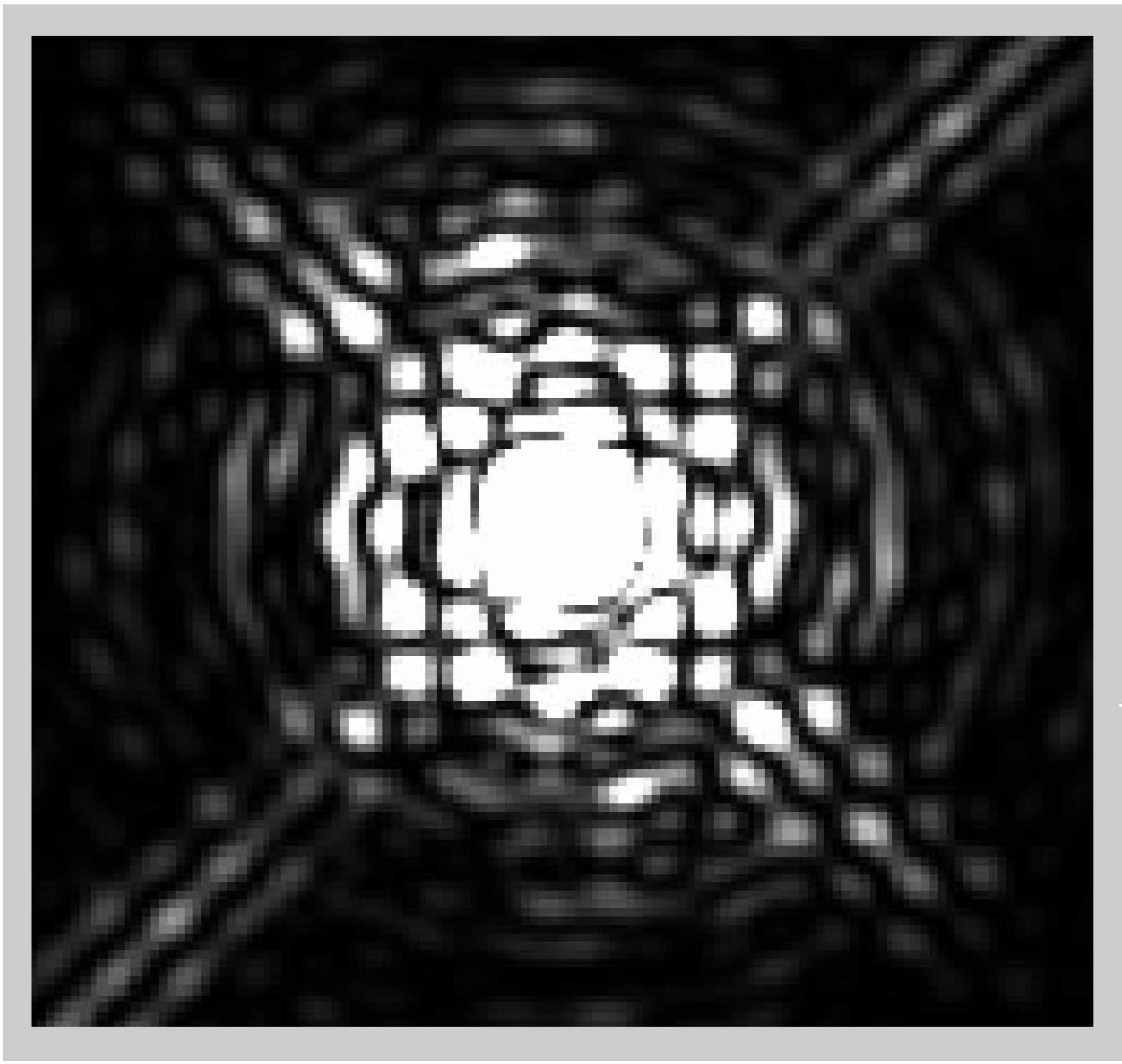}
\caption{PSF constructed by the Tiny Tim software for one of the frames in each set. We can easily see the spikes and the complex structure of the NIC--2 PSFs whatever the filter. {\it Left}: F160W, the grey scale goes from 0\% (black) to 0.13\% (white) of the peak intensity; {\it Right}: F180M, the grey scale goes from 0\% (black) to 0.16\% (white) of the peak intensity.}
\label{tinytim}
\end{figure}

\begin{enumerate}
\item First, for each individual image, we determine an improved PSF following the method described in Magain et al. (\cite{psfsimult}).  This is done by adding a numerical background to the approximate PSF, $s_0(\vec{x})$) (here, the deconvolved Tiny Tim PSF), so that the observed image $d_0(\vec{x})$ is better reproduced.  But, since this method assumes the image contains only point sources, and since our object contains a diffuse component, a part of it will be wrongly included in the improved PSF $s_1(\vec{x})$.  If the structure
of the diffuse component was identical around each point source, it would be entirely included in the PSF.  On the other hand, if it was completely different around each of the four point sources, only $\sim 25$\% of it would be included in the PSF.  In practice, a variable fraction of that diffuse component goes into the PSF.  As long as that fraction is below 100\%, our iterative procedure will allow to improve the results.
\item We then use that improved PSF $s_1(\vec{x})$ to perform a simultaneous deconvolution of all images ($s_1(\vec{x})$ slightly varies from image to image). This allows us to obtain a first approximation of the diffuse background $b_1(\vec{x})$. By construction, $b_1(\vec{x})$ is the same in all images.  However, since a part of the actual background has been included in the PSF $s_1(\vec{x})$, $b_1(\vec{x})$ is only the remaining part of the actual background.
\item We subtract $b_1(\vec{x})$, reconvolved and resampled to the initial resolution, from the original images. This gives us a new version of the observed images, $d_1(\vec{x})$, containing a lower amount of diffuse background.  The first iteration is over.
\item  To begin the second iteration, we use these images $d_1(\vec{x})$ to determine improved PSFs $s_2(\vec{x})$. As $d_1(\vec{x})$ contains a lower amount of background than $d_0(\vec{x})$, the new PSFs are indeed closer to the correct ones.
\item The simultaneous deconvolution of the original images $d_0(\vec{x})$ with the new PSFs $s_2(\vec{x})$ allows us to get a diffuse background $b_2(\vec{x})$ which is improved with respect to the one obtained at the previous iteration. 
\item We subtract $b_2(\vec{x})$ from the original images $d_0(\vec{x})$. This closes the second iteration.
\item The iterative procedure is continued until no significant improvement is observed. Usually around 4--5 iterations are necessary, depending on the structure under the sources.
\end{enumerate}

\section{Results}
\label{results}

\subsection{Iterative Process}

We now consider the application of this iterative process to the two sets of HST/NIC--2 images of the Cloverleaf.

\begin{figure} [b]
\centering
\includegraphics[width=4.2cm]{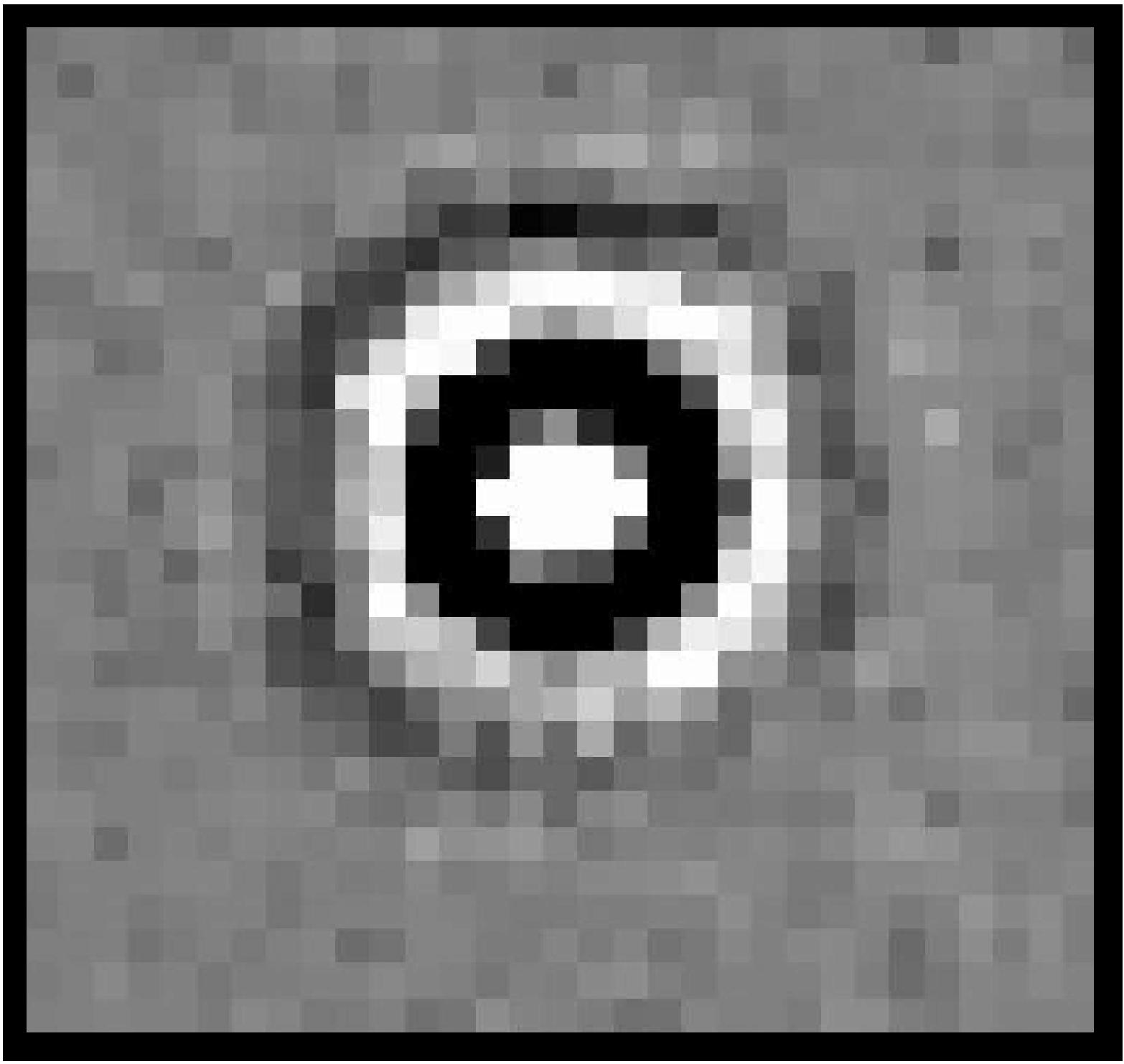}
\includegraphics[width=4.2cm]{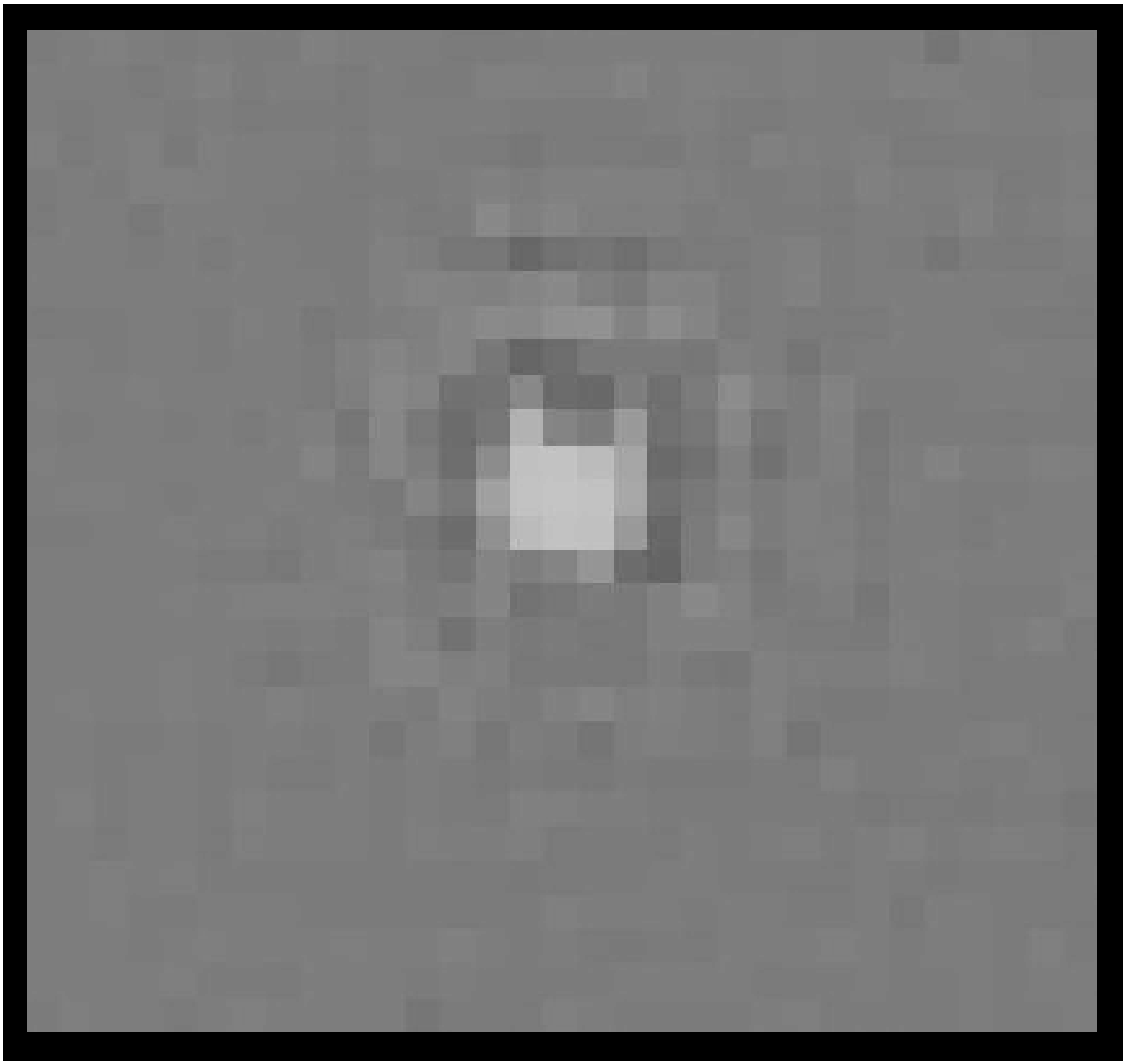}
\includegraphics[width=4.2cm]{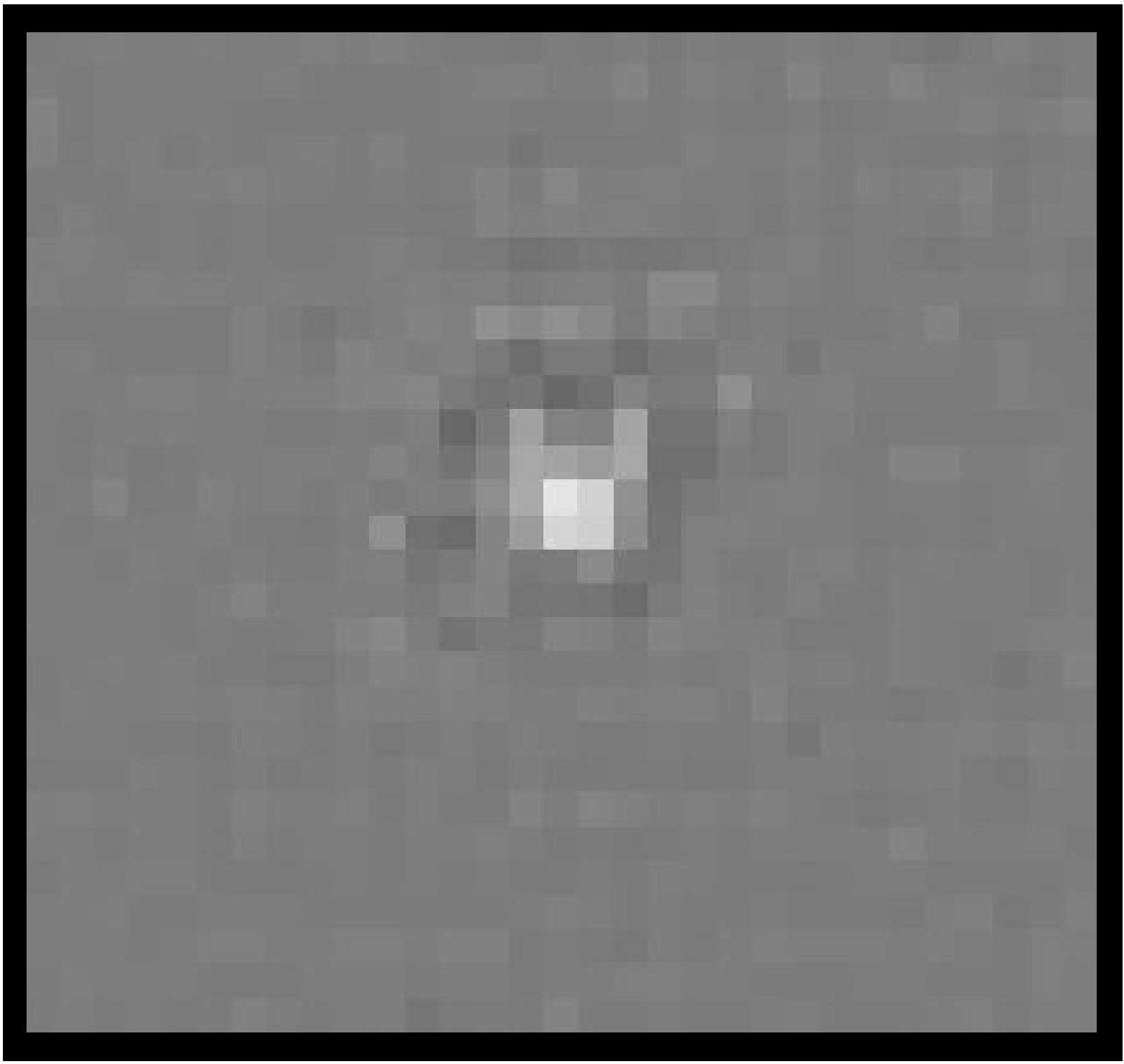}
\includegraphics[width=4.2cm]{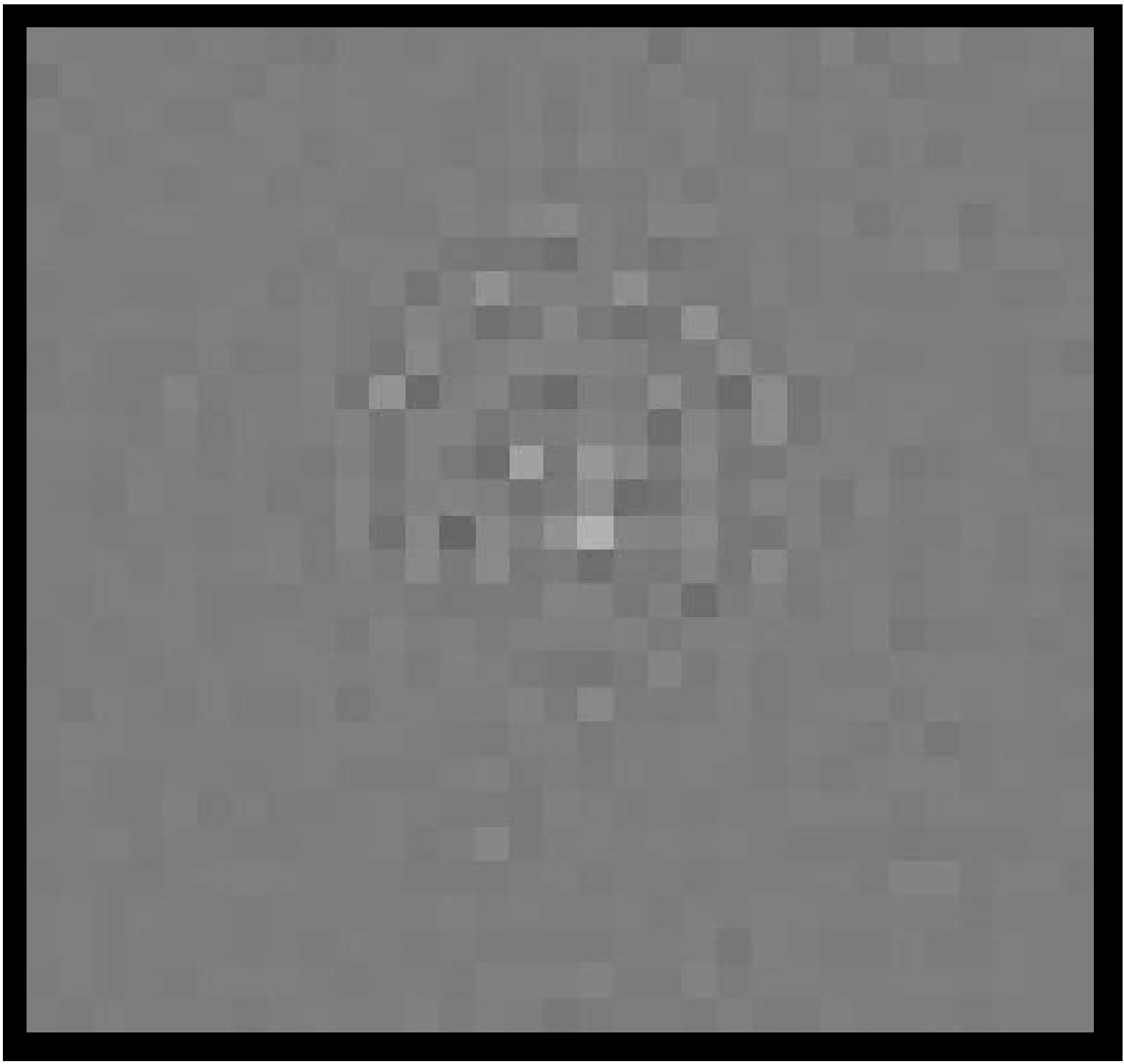}
\caption{Corrections applied to the PSFs at different stages of the process for one of the images of the F160W data set. The grey scale goes from -2.6\% (black) to +2.6\% (white) of the peak intensity of the deconvolved Tiny Tim PSF. {\it Top left} : corrections to the PSF in the first iteration (starting from the deconvolved Tiny Tim PSF). {\it Top right} : corrections at the second iteration. {\it Bottom left} : corrections at the fourth iteration. {\it Bottom right} : corrections at the last iteration.}
\label{psf_set1}
\end{figure}

\begin{figure}
\centering
\includegraphics[width=4.2cm]{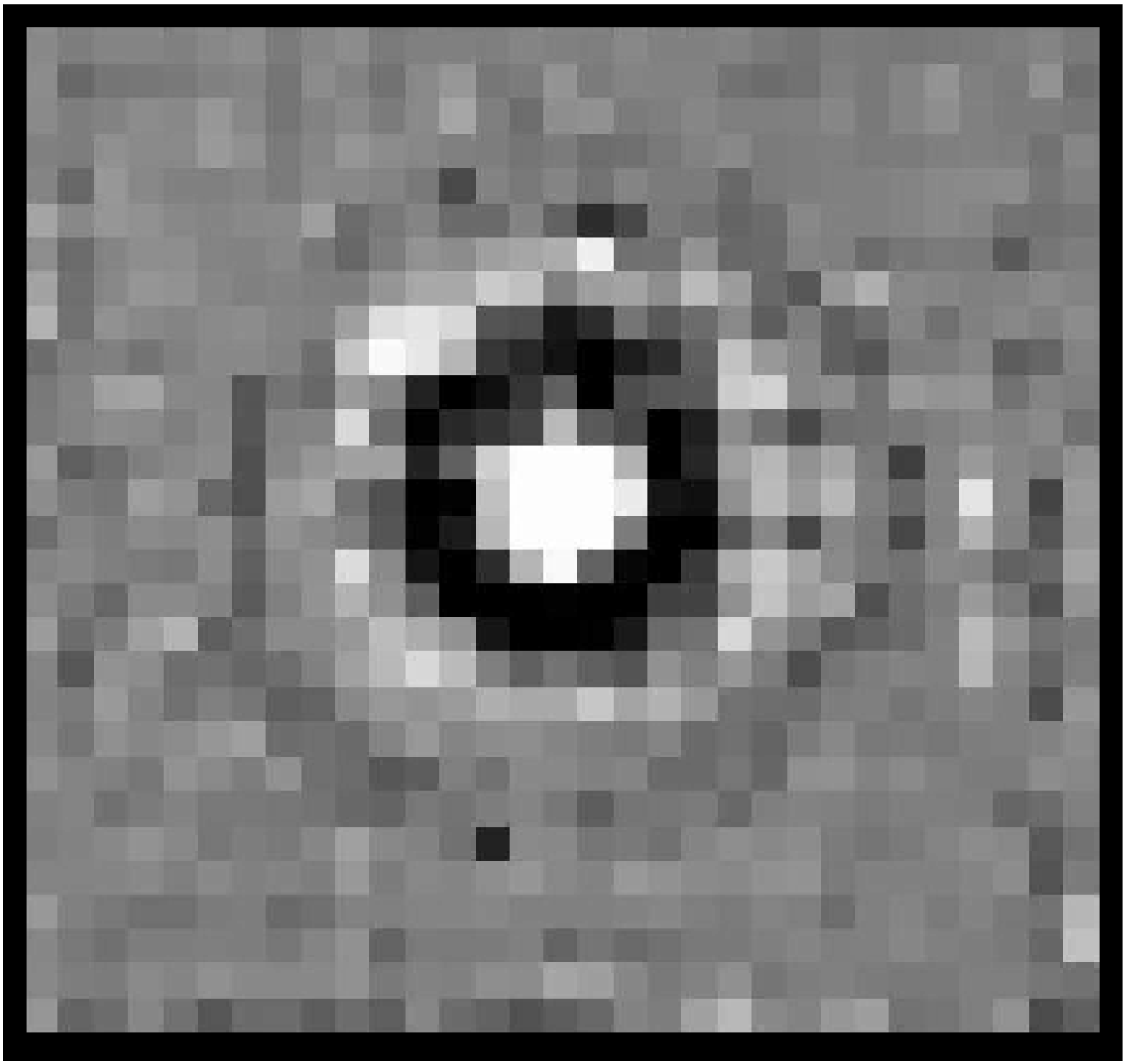}
\includegraphics[width=4.2cm]{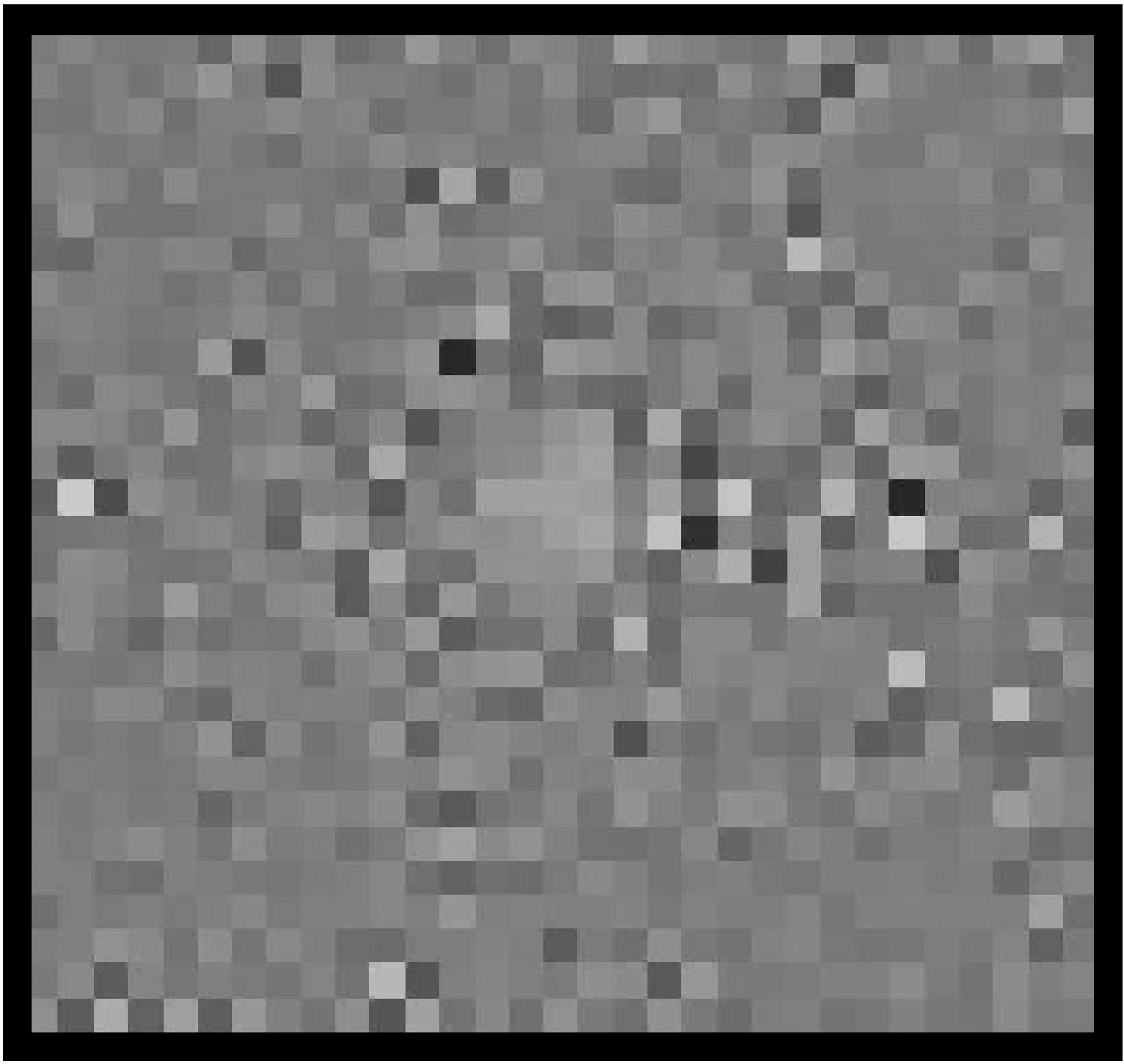}
\caption{Corrections applied to the PSFs at different stages of the process for one of the images of the F180M data set. The grey scale goes from -4.8\% (black) to +4.8\% (white) of the peak intensity of the deconvolved Tiny Tim PSF. {\it Left} : corrections to the PSF in the first iteration (starting from the deconvolved Tiny Tim PSF). {\it Right} : corrections at the last iteration}
\label{psf_set3}
\end{figure}

For the F160W data set, 7 iterations were necessary while, for the F180M data set, convergence was reached after 3 iterations.  This difference is due to the fact that the diffuse background is less intense relatively to the point sources in the latter filter. Figures \ref{psf_set1} and \ref{psf_set3} illustrate the evolution of the PSF in the iterative scheme : they show the corrections applied at different stages. We can see that the first step of the iterative process changes significantly the PSF obtained with Tiny Tim. The next steps allow to adjust smaller and smaller details. In the case of the F180M filter, it is obvious that only 3 iterations are necessary, as the corrections already become negligible after the second step. The same happens after the sixth iteration for the F160W data set.

\begin{figure}
\centering
\includegraphics[width=4.2cm]{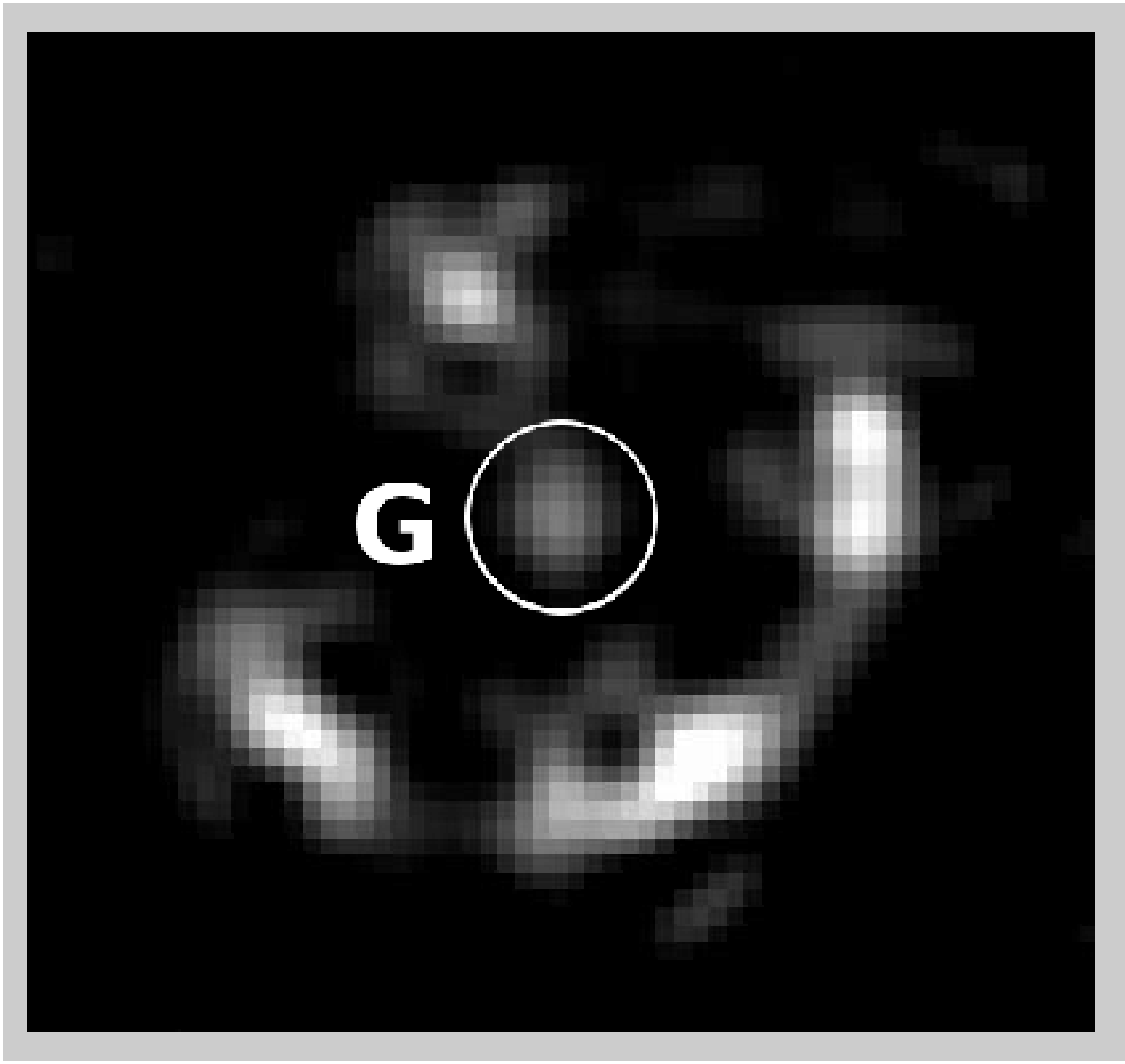}
\includegraphics[width=4.2cm]{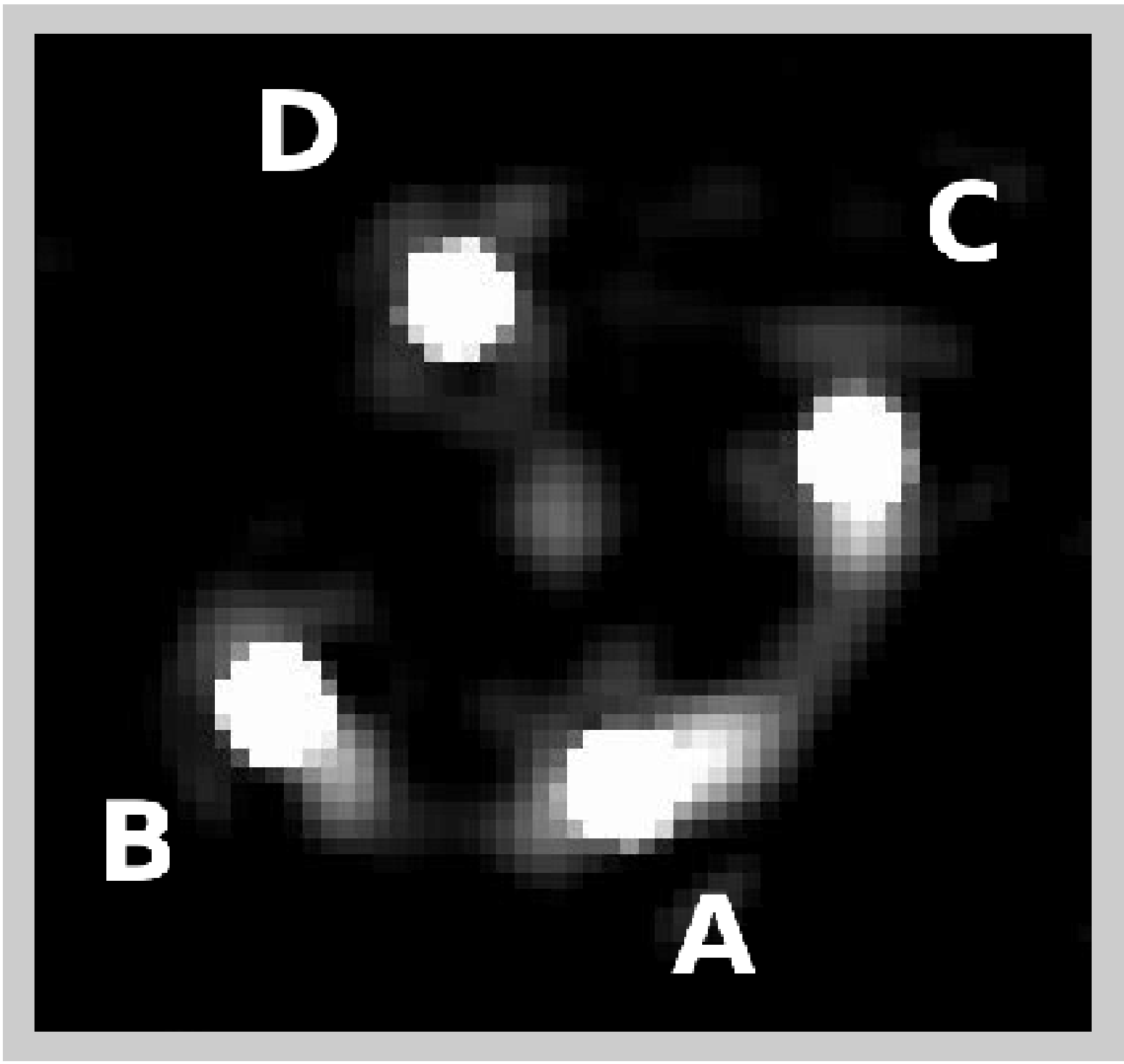}
\includegraphics[width=4.2cm]{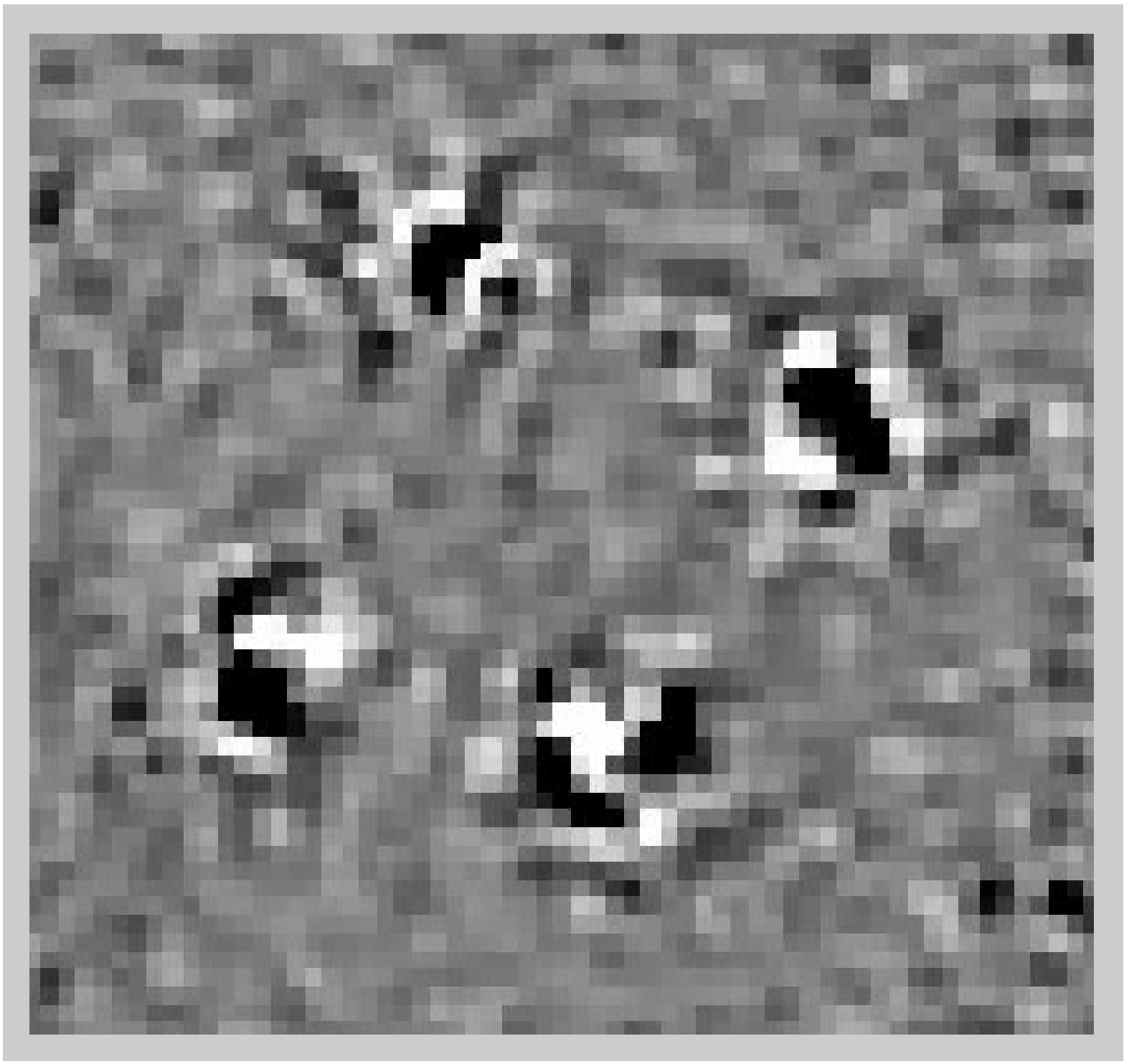}
\includegraphics[width=4.2cm]{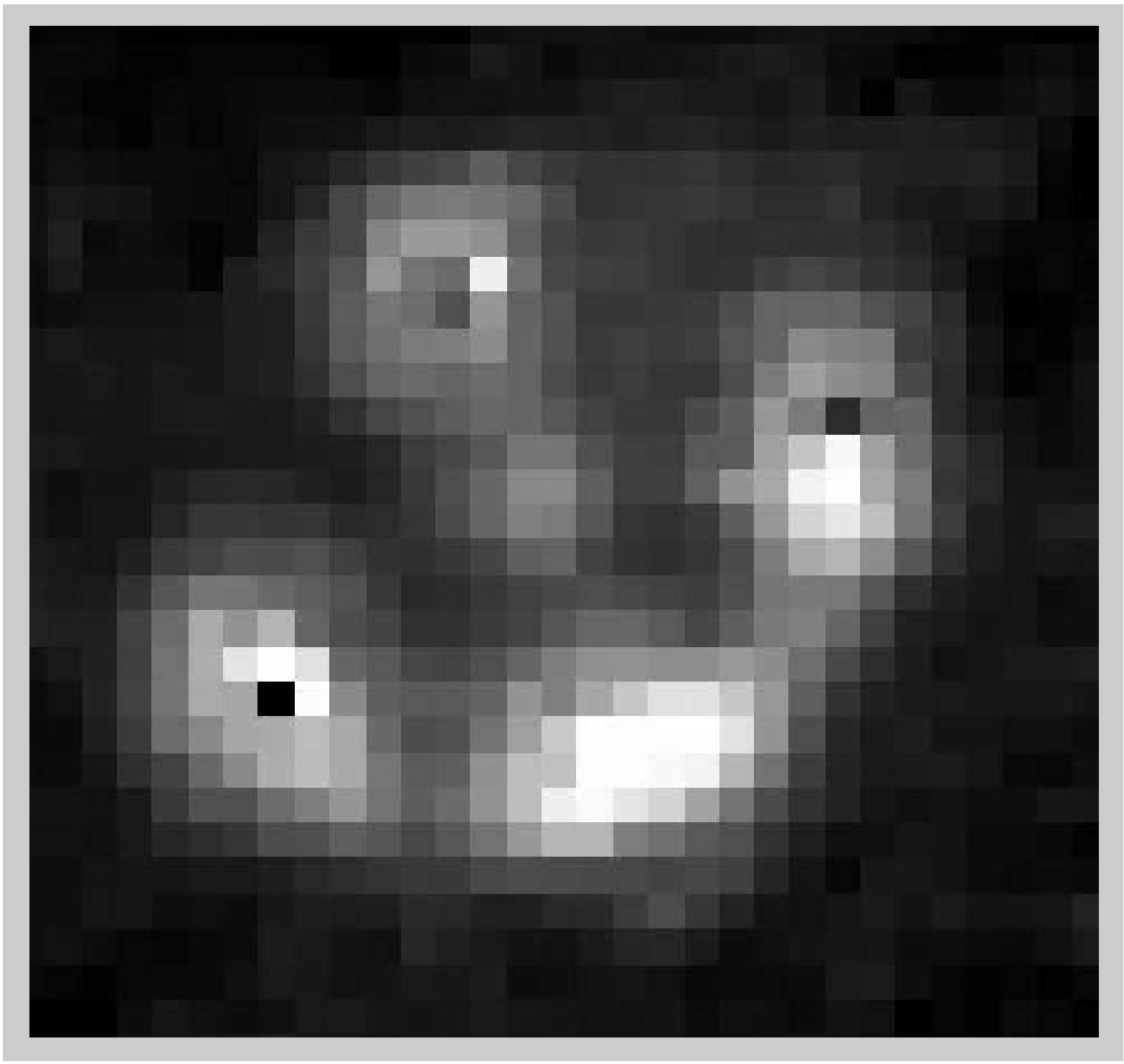}
\caption{Final results of the simultaneous deconvolution for the F160W data set. North is to the top and East to the left. {\it Top left} : smooth background common to all images of the set where the lensing galaxy is encircled. {\it Top right} : deconvolved image (point sources plus smooth background); the point sources are labelled as in Magain et al. (\cite{magain_a}). {\it Bottom left} : mean residual map of the simultaneous deconvolution. {\it Bottom right} : image reconvolved to the instrument resolution, with the point sources removed.}
\label{deconvo_set1}
\end{figure}

\begin{figure}
\centering
\includegraphics[width=4.2cm]{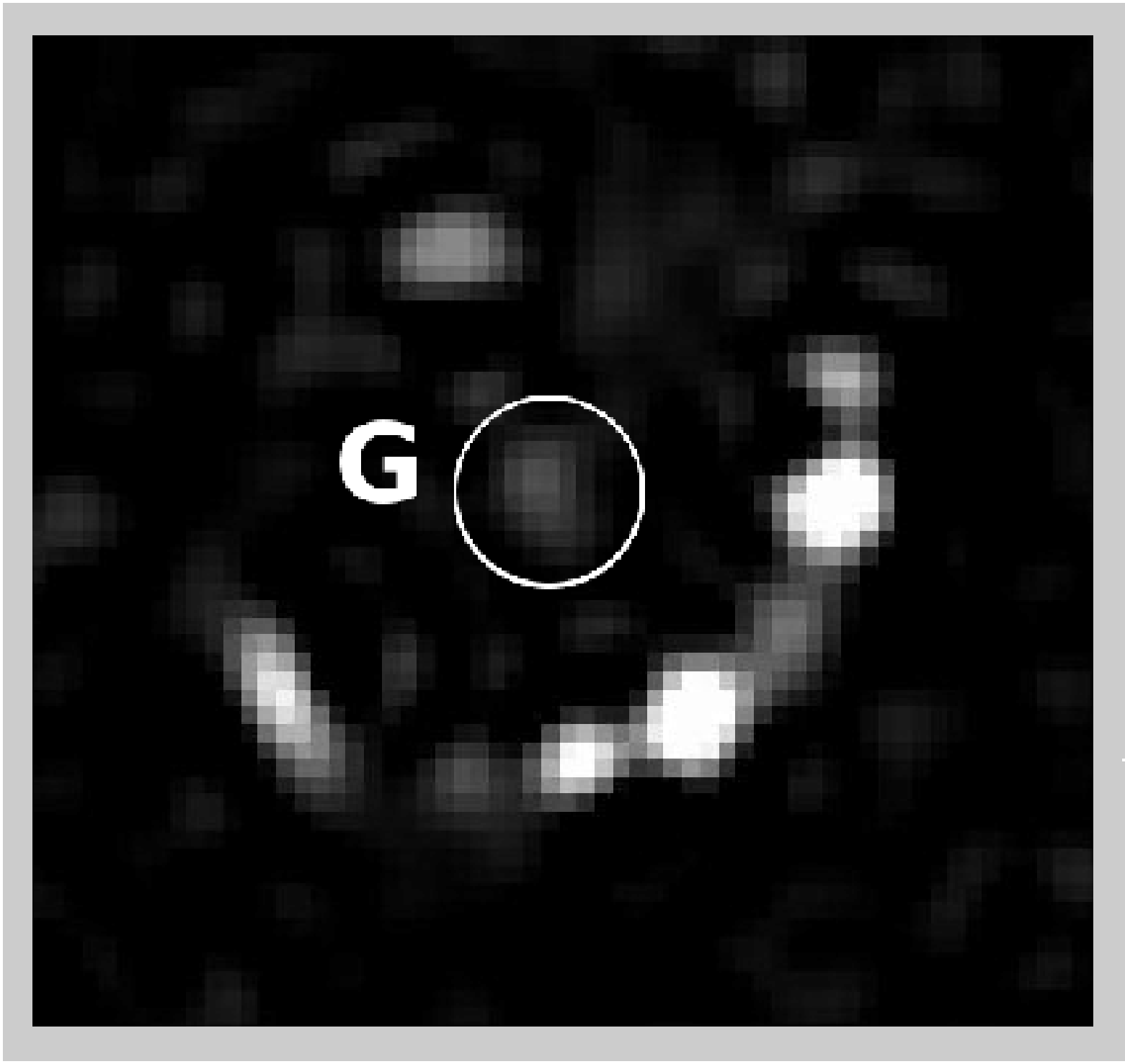}
\includegraphics[width=4.2cm]{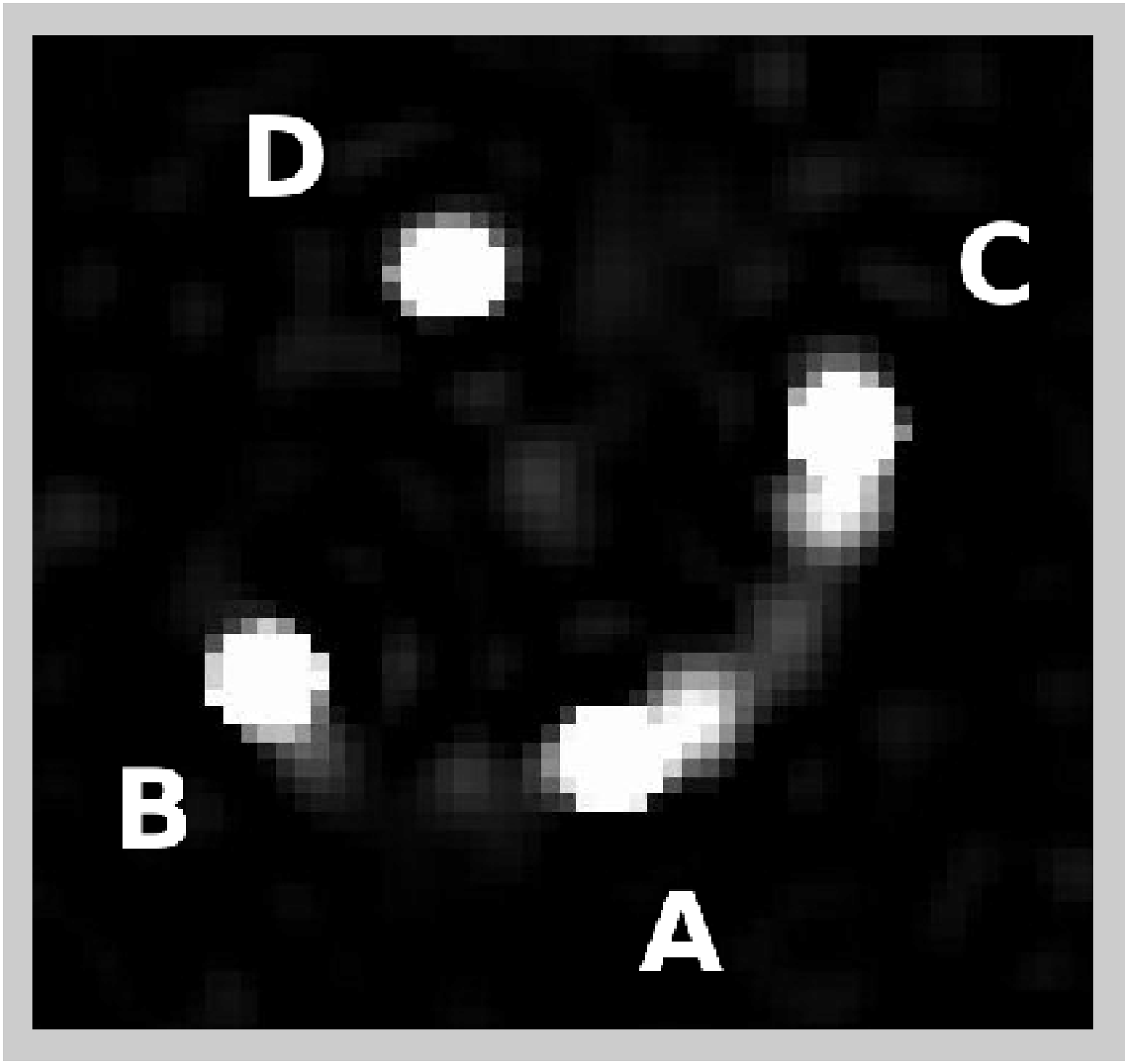}
\includegraphics[width=4.2cm]{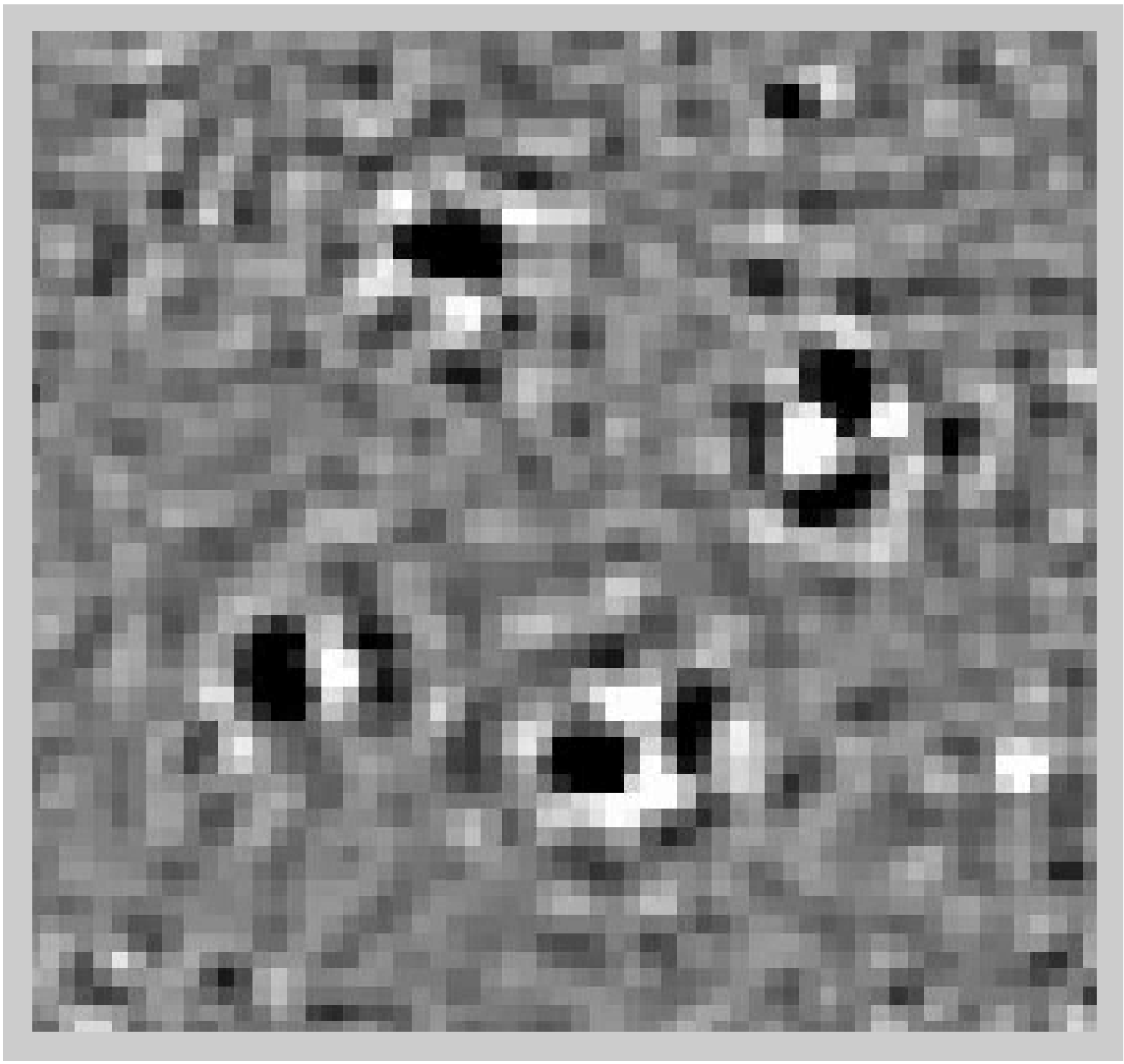}
\includegraphics[width=4.2cm]{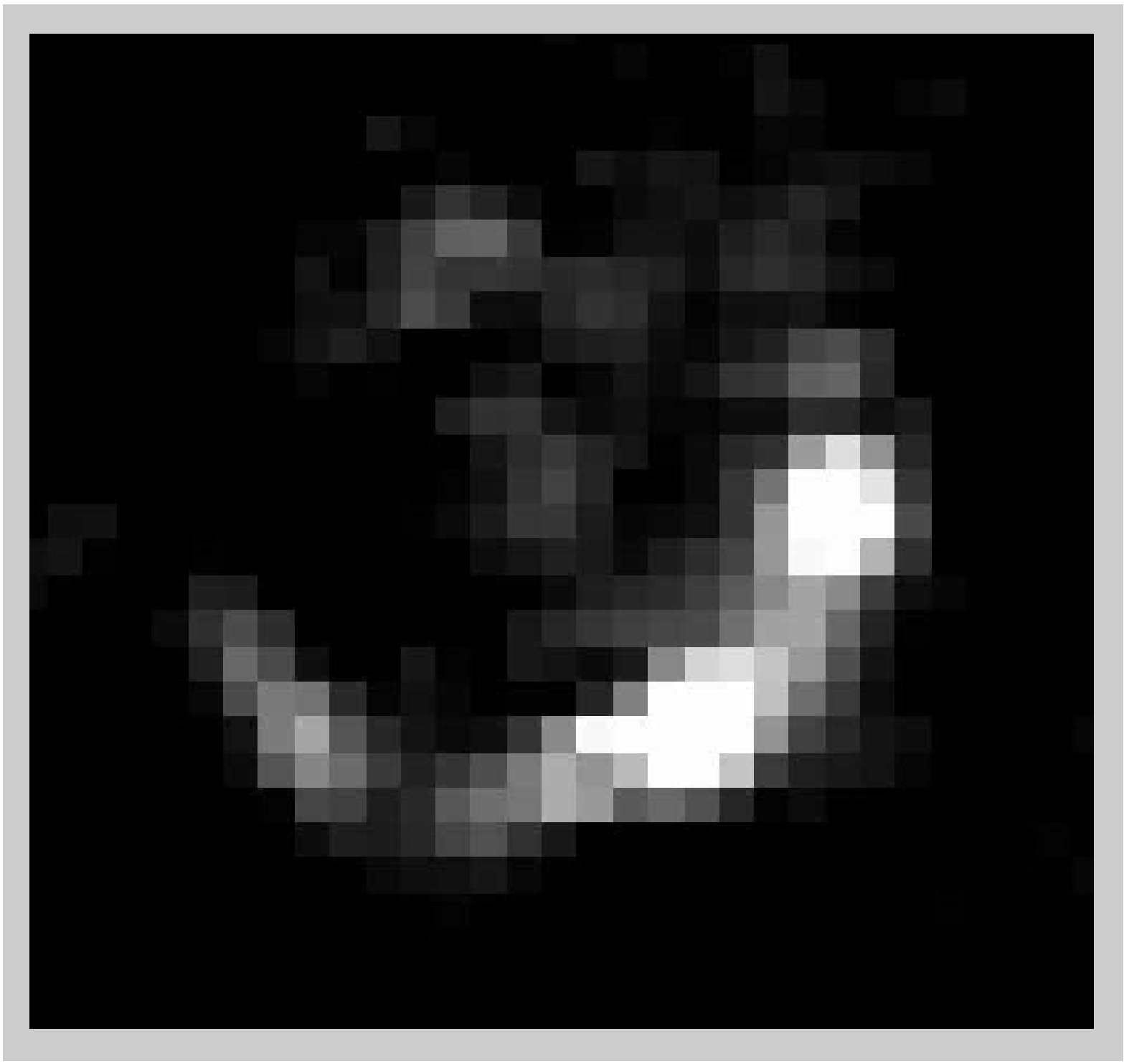}
\caption{Final results of the simultaneous deconvolution for the F180M data set. North is to the top and East to the left. {\it Top left} : smooth background common to all images of the set where the lensing galaxy is encircled. {\it Top right} : deconvolved image (point sources plus smooth background); the point sources are labelled as in Magain et al. (\cite{magain_a}). {\it Bottom left} :  mean residual map of the simultaneous deconvolution. {\it Bottom right} : image reconvolved to the instrument resolution, with the point sources removed.}
\label{deconvo_set3}
\end{figure}

Now that we have an idea about the evolution of the PSFs, we can focus on the results of the deconvolution itself. Figures \ref{deconvo_set1} and \ref{deconvo_set3} show the deconvolved frames from the last iteration, respectively for the F160W data set and for the F180M one. The partial Einstein ring, which is the gravitationally lensed image of the quasar host galaxy, and the lensing object can be seen for both sets on the background frame (top left) and on the background plus point sources frame (top right). The lens galaxy appears less intense compared to the point sources in the F180M filter, which is expected as this is a medium band filter including the  [O{\sc III}] emission lines (499--501 nm) at the redshift of the QSO and no expected emission line at the redshift of the lens. The partial Einstein ring also has a different structure: compared to the F160W filter, it appears more intense close to the point sources and less intense in between them.  This suggests that the narrow--line region (NLR) is more compact than the global lens galaxy, which could have been expected. An inversion of these light distributions could allow to reconstruct an image of the host galaxy and NLR.  This would be the first time one can map the host and NLR of a BAL QSO at such a high redshift.

The residuals $r_i$ from the deconvolution after the $i^{th}$ iteration are defined as follows : 
\begin{equation}
\label{resi}
\centering
 r_i(\vec{x}) \ = \ \frac{m_i(\vec{x}) \ - \ d_0(\vec{x})}{\sigma_0(\vec{x})} 
\end{equation}
where $m_i$ stands for the reconvolved model after the $i^{th}$ iteration, $d_0(\vec{x})$ for the observed image and $\sigma_0(\vec{x})$ for the standard deviation of this image. The residual map, as shown on the bottom left of Figs. \ref{deconvo_set1} and \ref{deconvo_set3}, is an important source of information : it guides us through the different steps. We can see there is some structure left under the point sources but nothing systematic and there is nearly no remnant structure where the ring and the lensing galaxy are located. The fact that the residuals under the four point sources have very different shapes suggests that they are not due to PSF errors, but rather to small PSF variations from one QSO image to another.

Another important guide through the different stages of the process is the reduced chi squared ($\chi^{2}_r$) which, theoretically, should be close to unity for a perfect deconvolution with a perfect PSF. In the last iterations it barely changes : the PSF is not improved significantly anymore and the iterative process has converged. We calculate it for each set and each iteration step in the zone of interest, i.e. in a square containing the four point sources and the extended structures (ring plus lens). We obtain a $\chi^{2}_r$ of 3.845 after the seventh iteration for the F160W data set, and a $\chi^{2}_r$ of $1.125$ for the F180M data set after the third iteration, which is really good. Let us mention that these values are computed taking into account all images of a given set, so that any slight incompatibility between some input images results in an increase of the $\chi^{2}$ that cannot be lowered by changing the model. A final $\chi^{2}_r$ of 1 means that the model is perfectly compatible with all the images of the set. It implies that all the images are perfectly compatible with each other and that the PSF is perfectly known. Any inaccuracy in the data acquisition or reduction will increase the final $\chi^{2}$.

\subsection{Astrometry and photometry}

Table \ref{astrophoto} gives the relative astrometry and photometry for the quasar images as well as for the lens galaxy in both filters. The coordinates are measured relative to component A (see Figs. \ref{deconvo_set1} and \ref{deconvo_set3}). The apparent magnitudes are given in the Vega system.

\begin{table*}
\centering 
\caption{Relative astrometric and photometric measurements for the four components of the system and the lensing galaxy (G). The right ascensions $\alpha$ and the declinations $\delta$ are given in arcsecond relative to component A. The photometry is given in apparent magnitudes in the Vega system. The internal $1 \ \sigma$ error bars are also indicated (see text for an explanation on how they are derived).}
\begin{tabular}{c||ccc|ccc}
\hline
\hline
 & \multicolumn{3}{c|}{F160W} & \multicolumn{3}{c}{F180M} \\
\hline
ID & $\Delta \alpha$ (\arcsec) & $\Delta \delta$ (\arcsec) & Magnitude & $\Delta \alpha$ (\arcsec) & $\Delta \delta$ (\arcsec) & Magnitude \\ 
\hline
A &  0.        & 0.        & 15.760 $\pm$ 0.002  
	& 0.         & 0.        & 15.548 $\pm$ 0.006 \\
B &  0.7426 $\pm$ 0.0002 & 0.1686 $\pm$ 0.0004 & 15.863 $\pm$ 0.005
	&  0.7458 $\pm$ 0.0003 & 0.1688 $\pm$ 0.0002 & 15.650 $\pm$ 0.009 \\
C & -0.4930 $\pm$ 0.0003 & 0.7135 $\pm$ 0.0004 & 16.143 $\pm$ 0.004
	& -0.4917 $\pm$ 0.0003 & 0.7105 $\pm$  0.0003 & 15.902 $\pm$ 0.004 \\
D &  0.3526 $\pm$ 0.0007 & 1.0394 $\pm$ 0.0004 & 16.400 $\pm$ 0.006 
	&  0.3532 $\pm$ 0.0003 & 1.0400 $\pm$ 0.0002 & 16.218 $\pm$ 0.007 \\
G &  0.1365 $\pm$ 0.0024 & 0.5887 $\pm$ 0.0035 & 20.527 $\pm$ 0.037
	&  0.1255 $\pm$ 0.0036 & 0.6192 $\pm$ 0.0069 & 22.182 $\pm$ 0.101 \\
\hline
\end{tabular}
\label{astrophoto}
\end{table*}

As the geometric distortions depend on the position on the detector, their proper corrections require an individual deconvolution of each image. We then obtain the position of each point source (relative to source A) on each deconvolved image, corrected for the distortion according to the formulae given in the NICMOS Data Handbook (Noll et al.,  \cite{Nicmos}) and compute average values. For the point sources, this gives more accurate results than a simultaneous deconvolution with a mean correction on the coordinates. On the other hand, this is not true for the lensing galaxy and Einstein ring.  As these are much fainter objects, it is better to rely on the results of the simultaneous deconvolution, where the signal in the whole set of images is used to constrain the shape of these objects. A mean geometric correction can then be applied, whose internal errors are lower than the random uncertainties on these fainter components.

The error bars given in Table \ref{astrophoto} are internal errors. They are calculated by deconvolving each image individually and comparing the coordinates and magnitudes obtained.  The listed values are the standard deviation of the means.

The astrometric precision for the point sources is about 0.5 milliarcsec in the F160W filter and 0.3 milliarcsec in the F180M filter. The higher precision in the medium band filter may be explained by the fact that the partial ring and the lens galaxy appear fainter relative to the point sources and thus have a lower contribution to the error bars.

Of course, the precision on the position of the lens galaxy is significantly lower.  This is due to the facts that (1) it is a diffuse object; (2) it is much fainter than the point sources (about 4.5 mag in the F160W filter and 6.4 mag in the F180M filter) and (3) it is mixed with the PSF wings of the point sources.

Table \ref{astrophoto} also shows that the results derived from both filters are not compatible within their internal error bars. As the geometry of the system is not expected to vary on the time scale of a few years, this disagreement suggests that the actual error bars are significantly larger than the internal errors. The causes may be diverse. As the two sets of data have been acquired 6 years apart, with a different orientation of the HST and thus of the detector, and in different cycles of NICMOS (pre-- and post--NCS, NICMOS Cooling System), some geometrical distortions may not have been completely taken into account. The uncertainties on the coefficients of the formulae used to correct for the geometrical distortions, as given in the NICMOS Data Handbook (Noll et al.\  \cite{Nicmos}) account for an uncertainty of the order of 0.1 milliarcsec in each filter, which is about an order of magnitude smaller than the external errors we obtain. It is thus possible that a residual distortion of the NICMOS images remains, at the $10^{-3}$ level (0.001 arcsec per arcsec). An imperfect separation of the partial Einstein ring from the point sources in the deconvolution process as well as some inaccuracies in the PSF recovery may also play a role.

The external errors, computed by comparing the source positions derived from the two data sets, are the following: the average difference between the point source positions amounts to 1.4 milliarcsec.  Assuming that the errors in both data sets equally contribute to this difference, we derive a value of $1.4/\sqrt{2} \ \approx 1 $ milliarcsec (i.e. 0.013 pixel) for the estimated accuracy in the position of the point sources.

Our measurements are compared to those of Magain et al. (\cite{magain_a}) and Turnshek et al. (\cite{turnshek}) in Table \ref{comp_astro}.  The latter were derived from images acquired with another HST intrument (Wide Field Planetary Camera) and with a completely different image processing technique, while the first ones were obtained from much lower resolution ground--based images. For both sets of results we indicate the $1 \ \sigma$ error bars (which do not appear in the original paper of Magain et al.).  The average difference between our results and those of Magain et al.\  (\cite{magain_a}) amount to 4 milliarcsec, which is about the error bars on the measurements performed by these authors.  The same comparison with Turnshek et al. (\cite{turnshek}) gives an average difference of 2.6 milliarcsec, also compatible with their error bars.

\begin{table*}
\centering 
\caption{Relative astrometry of the Cloverleaf from Magain et al. (\cite{magain_a}) and from Turnshek et al. (\cite{turnshek}). The right ascension $\alpha$ and the declination $\delta$ are given in arcsecond relatively to component A. The $1 \ \sigma$ error bars are also indicated.}
\begin{tabular}{c||cc|cc}
\hline
\hline
& \multicolumn{2}{c|}{Magain et al. (\cite{magain_a})} & \multicolumn{2}{c}{Turnshek et al. (\cite{turnshek})} \\
\hline
ID & $\Delta \alpha$ (\arcsec) & $\Delta \delta$ (\arcsec) & $\Delta \alpha$ (\arcsec) & $\Delta \delta$ (\arcsec) \\ 
\hline
A &  0.        & 0.        
	& 0.         & 0.         \\
B &  0.753  $\pm$ 0.006 & 0.173 $\pm$0.006
	&  0.744 $\pm$ 0.003 & 0.172 $\pm$ 0.003  \\
C & -0.496 $\pm$ 0.004 & 0.713 $\pm$ 0.003
	& -0.491 $\pm$ 0.003 & 0.716 $\pm$  0.004 \\
D &  0.354 $\pm$ 0.004 & 1.043 $\pm$ 0.004
	&  0.355 $\pm$ 0.003 & 1.043 $\pm$ 0.012 \\
\hline
\end{tabular}
\label{comp_astro}
\end{table*}

The primary lens, a single galaxy, was detected in \cite{kneib} by Kneib et al. After a PSF subtraction of the four lensed images they obtained the following relative positions for the lensing galaxy :
\begin{center}
$\alpha \ = \ 0.112 \arcsec \ \pm \ 0.02 \arcsec$ \\
\bigskip
and \\
\bigskip
$\delta \ = \ 0.503 \arcsec \ \pm \ 0.02 \arcsec$ \\
\end{center}
Their lens position is compatible with our result in right ascension ($\Delta \alpha = -0.025 \pm 0.020$). But this is not true for the declination ($\Delta \delta = -0.086 \pm 0.020$). Possible systematic errors, in particular on the lens position, are investigated in the next section.

\section{Synthetic Image}
\label{synthetic}

\begin{figure} [h]
\centering
\includegraphics[scale=0.24]{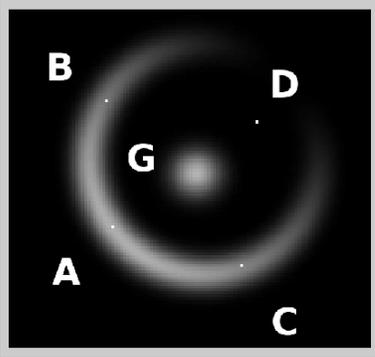}
\caption{Synthetic image of a gravitationally lensed quasar with a configuration similar to the Cloverleaf : 4 point sources, a faint lensing object and a partial Einstein ring. The orientation is the same as the original F160W Cloverleaf images.}
\label{arti}
\end{figure}

\begin{figure}
\centering
\includegraphics[scale=0.26]{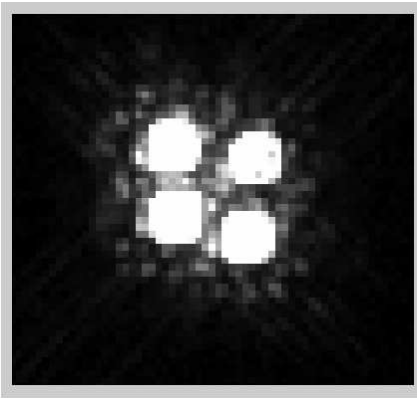}
\caption{Synthetic image convolved with a HST-type PSF unknown to the test performer and with added random noise similar to the actual observation.}
\label{articonv}
\end{figure}

The accuracy of our results is further tested by carrying the same procedure on a synthetic image having characteristics similar to those of the HST/NICMOS F160W Cloverleaf image : 4 point sources, a faint lensing object and a partial Einstein ring (see Fig.\  \ref{arti}). This synthetic image was convolved  with a PSF similar to the actual one but unknown to the test performer. Random noise was then added to get a S/N comparable to that of the combined HST image (see Fig.\  \ref{articonv}).

\begin{figure}
\centering
\includegraphics[scale=0.26]{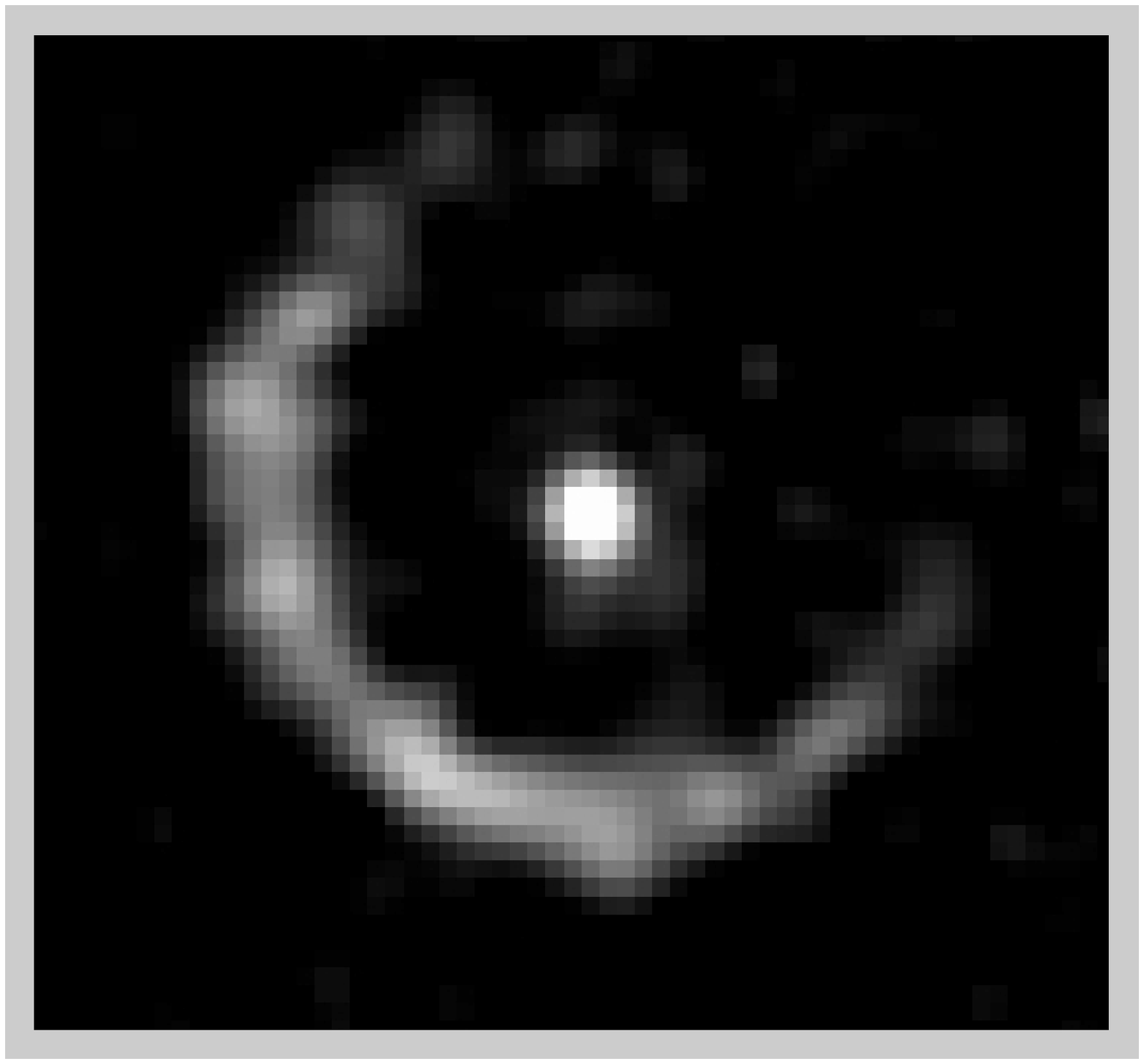}
\includegraphics[scale=0.26]{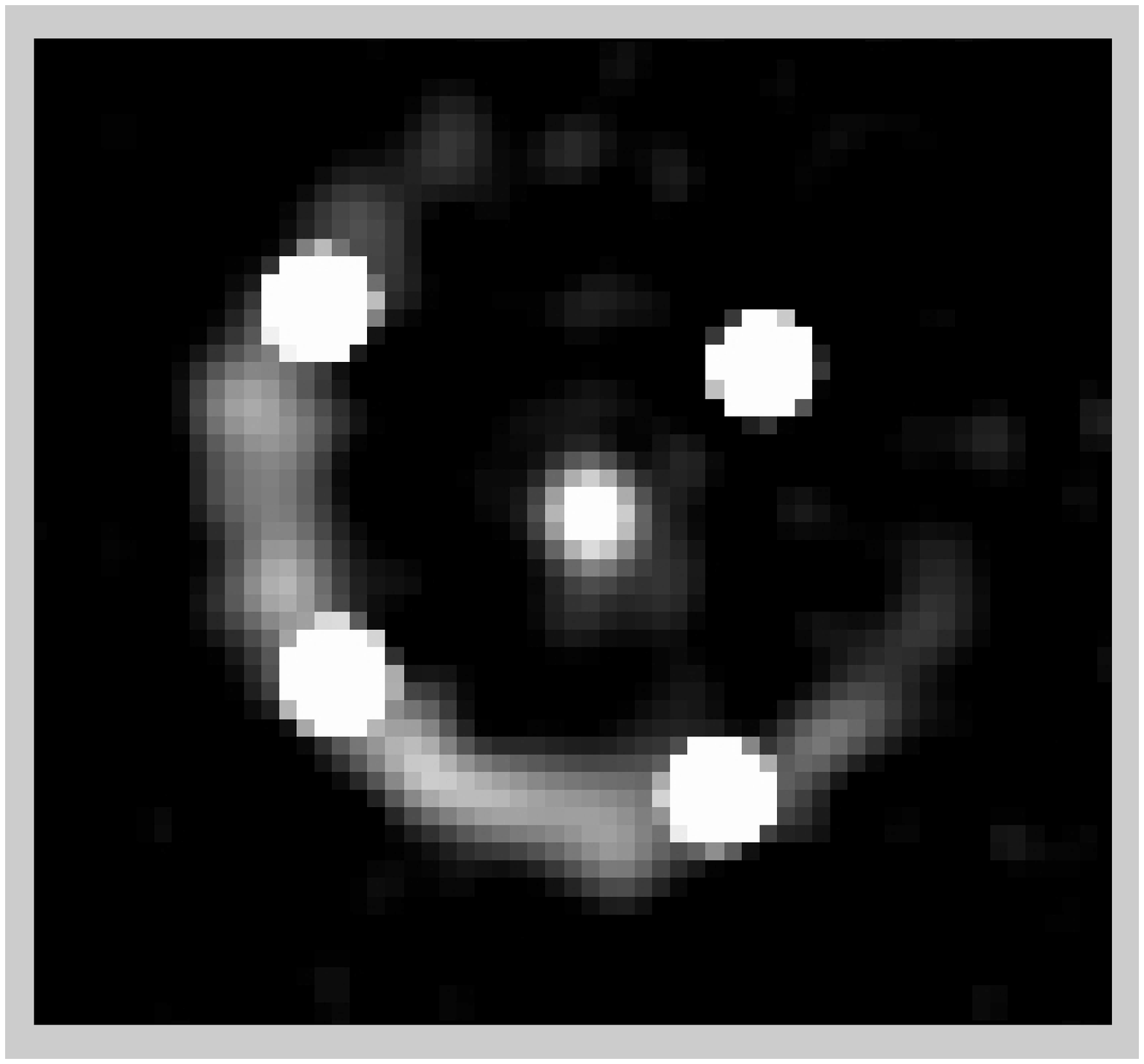}
\includegraphics[scale=0.26]{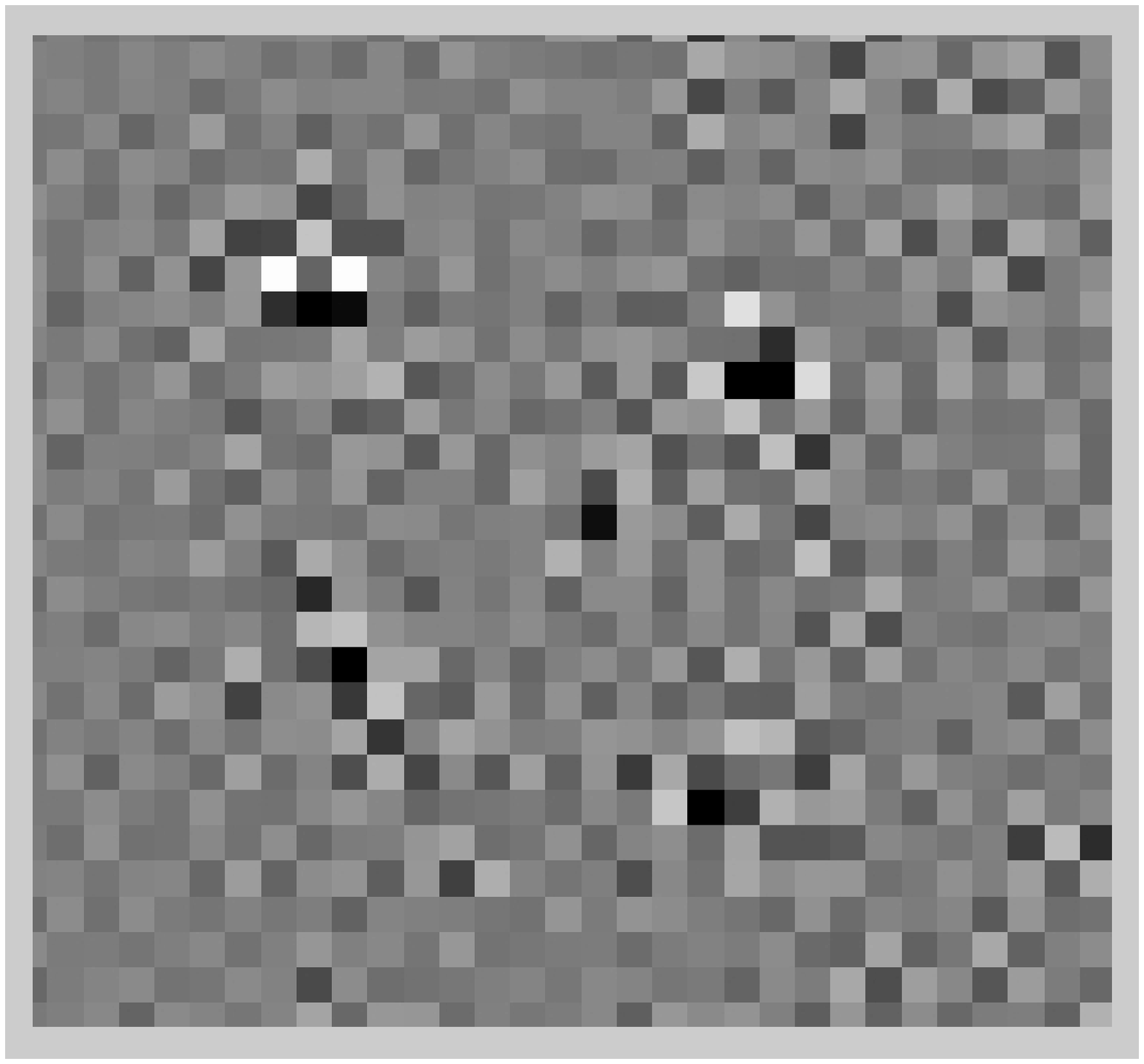}
\caption{Results of the last iteration on the synthetic image. {\it Top} : diffuse background.  {\it Middle} : diffuse background plus point sources. {\it Bottom} : residual map of the deconvolution. }
\label{artires}
\end{figure}

The results obtained after three iterations are presented on Fig.\  \ref{artires}, which displays the background alone, the point sources plus background and finally the residual map.  Some remnant structures can be seen under the point sources on the residual map. They are slightly weaker than those observed in the residual maps of the actual images, but show similar characteristics.

On average, the flux in the background (ring + lens) is recovered within 4\%, which can be considered as excellent since this diffuse background is very weak compared to the point sources.  However, because of the smoothing constraint, the deconvolved ring and lens appear slightly smoother than the original ones.  The largest differences are found under the brightest point source (A), where the deconvolved ring is about 43\% below the original one.

Table \ref{arti_astrom} summarizes the astrometry carried out on this artificial Cloverleaf : the first pair of columns present the measurements made on the final deconvolved image resulting from the iterative process, the second pair of columns the results when using a deconvolved Tiny Tim PSF for a unique deconvolution and the last one the measurements made on the original image.

The differences between the positions obtained for a particular source reach a maximum of about 0.3 milliarcsec with a mean value around 0.1 milliarcsec, which is slightly better than the internal precision estimated in Table \ref{astrophoto}.  On the other hand, the lens galaxy position is not as accurate : the maximum difference amounts to 20 milliarcsec (i.e.\  a quarter of a pixel). Indeed, the position of such very faint diffuse objects is rather sensitive to inaccuracies in the PSF: any error in the wings of the bright point source PSFs may have impacts on the faint neighbouring objects.

Given these possible sources of errors and the results of the simulations, we estimate the accuracy on the lens galaxy position to amount to some 20 milliarcsec.

\begin{table*}
\centering 
\caption{Relative astrometry of the artificial Cloverleaf. The two coordinates are given in arcsecond relatively to component A.}
\begin{tabular}{c||cc|cc|cc}
\hline
\hline
& \multicolumn{2}{c|}{Iterative process} & \multicolumn{2}{c|}{Tiny Tim} & \multicolumn{2}{c}{Original image}\\
\hline
ID & $\Delta \alpha$ (\arcsec) & $\Delta \delta$ (\arcsec) & $\Delta \alpha$ (\arcsec) & $\Delta \delta$ (\arcsec) & $\Delta \alpha$ (\arcsec) & $\Delta \delta$ (\arcsec) \\ 
\hline
A &  0.    & 0.     	
	&  0.    & 0.    
	&  0.    & 0.\\
B &  0.7719 & 0.1770  	
	&  0.7726 & 0.1755 
	&  0.7718 & 0.1767\\
C & -0.4538 & 0.7140  
	& -0.4538  & 0.7129 
	& -0.4538 & 0.7138\\
D &  0.3913 & 1.0480  
	&  0.3921 & 1.0483 
	&  0.3913 & 1.0479\\
G &  0.1826 & 0.6151 
	&  0.1787 & 0.5877 
	&  0.1819 & 0.5940\\
\hline
\end{tabular}
\label{arti_astrom}
\end{table*}

\section{Conclusions}
\label{conclusion}

We have elaborated a new image processing method, based on the MCS deconvolution algorithm, which allows, at the same time, to determine the Point Spread Function and to deconvolve a set of images. It is applicable to images which contain at least 2 point sources so that the algorithm can separate the contributions of background objects from those of the PSF itself.

This technique is particularly well suited to the analysis of multiply imaged quasars: it allows to separate extended structures (lensing galaxy, arcs or rings) from the point sources. It provides accurate photometry and astrometry, which is very important for modelling the lensed systems.

Our internal error bars on the source positions, taking into account the error coming from the deconvolution only, are of the order of 0.4 milliarcsec. When comparing the astrometry coming from two different sets of images, we find external errors of the order of 1 milliarcsec. They probably find their origin in an incomplete correction of the geometric distortions.

Moreover, we detect the lensing galaxy and measure its position with an accuracy of the order of 20 milliarcsec, and discover a partial Einstein ring, which should allow to constrain the deflection model and, through inversion of the lens equations, to estimate the light distribution in the quasar host galaxy and narrow line region.

\begin{acknowledgements}
The authors would like to thank Sandrine Sohy for her help and commitment in the programing part of the work.  This work has been supported by ESA and the Belgian Federal Science Policy Office under contract PRODEX 90195.
\end{acknowledgements}


\begin{thebibliography}{}

\bibitem[1998]{courbin}
Courbin F., Lidman C., Magain P., 1998, A\&A, 330, 57

\bibitem[1984]{hazard}
Hazard C., Morton D.C., Terlevich R. \& McMahon R., 1984, ApJ, 282, 33

\bibitem[1998]{kneib}
Kneib J.--P., Alloin D. \& Pell\'o R., 1998, A\&A, 339, 65

\bibitem[2004]{tinytim}
Kris J. \& Hook R., 2004, http://www.stsci.edu/software/tinytim

\bibitem[1988]{magain_a} 
Magain P., Surdej J., Swings J.--P., Borgeest U., Kayser R., Kühr H., Refsdal S. \& Remy M., 1988, Nature, 334, 6180

\bibitem[1998]{MCS}
Magain P., Courbin F. \& Sohy S., 1998, ApJ, 494, 472

\bibitem[2006]{psfsimult}
Magain P., Courbin F., Gillon M., Sohy S., Letawe G., Chantry V. \& Letawe Y., 2006, A\&A, in press (astro-ph/0609600)

\bibitem[2004]{Nicmos}
Noll, K., et al. 2004, "NICMOS Instrument Handbook", Version 7.0, Baltimore, STScI


\bibitem[1997]{turnshek}
Turnshek D. A., Lupie O. L., Rao S. M., Espey B. R. \& Sirola C. J., 1997, ApJ, 485, L100--111

\end{thebibliography}
\end{document}